\documentclass[a4paper, 10pt]{article} 

\usepackage[margin=2.54cm]{geometry}

\usepackage{ccaption}
\usepackage{float}
\usepackage{wrapfig}
\usepackage{appendix}
\usepackage{cite}
\usepackage{subfigure}
\usepackage{adjustbox}
\usepackage{nicematrix}
\usepackage{float}
\usepackage{subcaption}
\usepackage{booktabs}

\usepackage{xcolor}
\usepackage{dsfont}
\usepackage[pdftex,colorlinks=true,linkcolor=blue,citecolor=blue,urlcolor=black]{hyperref}

\usepackage{amssymb}
\usepackage{amsmath}
\usepackage{amsthm}
\usepackage{mathtools}
\usepackage{physics}
\usepackage{braket}
\usepackage{enumitem} 
\usepackage{authblk}
\usepackage{stix2}

\DeclarePairedDelimiter{\kett}{\lVert}{\rAngle}
\DeclarePairedDelimiterX\bbrakett[2]{\lAngle}{\rAngle}{#1 \Vert #2}

\newsavebox{\mstrut}
\newcommand{\ex}[1]{\langle #1 \rangle}

\newcommand{\mF}[0]{\mathcal{F}}
\newcommand{\mS}[0]{\mathcal{S}}

\newcommand{\ndp}[0]{\mathcal{N}_{\text{dep}}(\lambda)}
\newcommand{\nmg}[0]{\mathcal{N}_{\text{mg-dep}}(\lambda)}
\newcommand{\ncat}[0]{\mathcal{N}_{\text{cat}}(\lambda)}
\newcommand{\ad}[0]{a^\dagger}

\newcommand{\email}[1]{\href{mailto: #1}{#1}}

\title{Non-zero noise extrapolation: accurately simulating noisy quantum circuits with tensor networks}

\author[1, 2, 3]{Anthony P. Thompson\thanks{\email{anthony.peter.thompson@pm.me}}}
\author[1, 4]{Arie Soeteman}
\author[1]{Chris Cade\thanks{\email{chris@fermioniq.com}}}
\author[1]{Ido Niesen}
\affil[1]{\small Fermioniq B.V., Amsterdam, NL}
\affil[2]{School of Mathematics, 
University of Bristol, UK}
\affil[3]{Quantum Engineering Center for Doctoral Training, University of Bristol, UK}
\affil[4]{Institute for Logic, Language and Computation, 
University of Amsterdam, NL}

\date{\today}

\begin{document}
\maketitle

\begin{abstract}
 Understanding the effects of noise on quantum computations is fundamental to the development of quantum hardware and quantum algorithms. Simulation tools are essential for quantitatively modelling these effects, yet unless artificial restrictions are placed on the circuit or noise model, accurately modelling noisy quantum computations is an extremely challenging task due to unfavourable scaling of required computational resources. Tensor network methods offer a scalable solution for simulating computations that generate limited entanglement or that have noise models which yield low gate fidelities. However, in the most interesting regime of entangling circuits (with high gate fidelities), relevant for error correction and mitigation, tensor network simulations often achieve poor accuracy.
 
In this work we develop and numerically test a method for significantly improving the accuracy of tensor network simulations of noisy quantum circuits in the low-noise (i.e. high gate-fidelity) regime. Our method comes with the advantages that it (i) allows for the simulation of quantum circuits under generic types of noise model, (ii) is especially tailored to the low-noise regime, and (iii) retains the benefits of tensor network scaling, enabling efficient simulations of large numbers of qubits. We build upon the observations that adding extra noise to a quantum circuit makes it easier to simulate with tensor networks, and that the results can later be reliably extrapolated back to the low-noise regime of interest. These observations form the basis for a novel simulation technique that we call non-zero noise extrapolation, in analogy to the quantum error mitigation technique of zero-noise extrapolation. 
\end{abstract}

\maketitle

\section{Introduction}\label{sec:intro}
Scalable, fault tolerant quantum computing holds promise to revolutionise computing across many scientific, industrial, and commercial domains. However, achieving this promise remains a major outstanding scientific and engineering challenge. Arguably the most significant obstacle is that the quantum information encoded in quantum hardware is susceptible to \emph{noise}: unwanted side effects that occur during computation. As such, there is a concentration of research efforts towards developing practical methods for detecting, mitigating, and correcting errors arising due to noise in quantum computers. Crucial to the development of such methods is a thorough understanding of the origins and effects of noise, which forms the basis for device-specific error mitigation techniques, noise-resilient quantum algorithms, more performant error-correcting codes, and more efficient decoders. 

In all of these endeavours digital simulations play a crucial role~\cite{ryan2021realization, fowler2012surface, bausch2024learning, 
marshall2023incoherent, darmawan2017tensor, google2023suppressing}. This can range from simulating the physics that governs the behaviour of the device at the lowest level all the way to emulations and simulations of quantum circuits at the gate level. The low-level simulations can be used to inform the gate-level simulations, which can in turn be used to study and develop error correction and mitigation strategies~\cite{PhysRevA.109.042620, Blume_Kohout_2017, Dahlhauser_2021, Mangini_2024, PhysRevLett.120.050505}. Two particularly important emerging applications of quantum circuit simulation include the estimation of threshold rates for error correcting codes, given a particular noise model and decoder~\cite{fowler2012surface,landahl2011fault,tomita2014low,heim2016optimal}, and the development of noise-resilient quantum algorithms~\cite{chen2024noiseawaredistributedquantumapproximate, PRXQuantum.2.010324}.

However, simulating noisy quantum circuits is a tremendously difficult task. Without specific restrictions on the circuit and noise model, the most widely used method for noisy circuit simulation is the \textit{full state} approach -- here, either the entire density matrix of the noisy state is stored in memory at all times, or the entire statevector of one `trajectory' through the noisy computation is stored. In the former, the memory requirements scale as $4^{n}$ for $n$ qubits, limiting such simulations to at most 15-20 qubits. In the latter, the memory requirements scale as $2^n$, limiting simulations to 30-40 qubits, at the expense of possibly requiring very many samples over trajectories. The scaling of such methods clearly limits their utility, especially in a time when quantum hardware devices are reaching scales of 50-100+ qubits.\footnote{For instance, simulating even the distance-5 Surface-49 QEC code requires the simulation of 25 noisy qubits, something that is out of reach of full density matrix simulators, and already challenging for statevector trajectory methods~\cite{o2017density,marshall2023incoherent}.}

As a consequence, circuits and noise models are often simplified to maintain tractability of the simulation. For example, circuits and noise models composed of Clifford gates can be efficiently simulated, and the assumption of Clifford noise is therefore commonly made in the error correction literature, often for the purpose of improving simulation speed and scalability~\cite{Flammia_2020}. However, such simplifications can limit the ability to understand and accurately predict the effects of noise on circuits running on real hardware subject to complex noise~\cite{o2017density}. There is therefore a need for scalable methods for emulating or simulating \textit{generic} noise on system sizes larger than what is feasible via full-state approaches.

Tensor networks offer an approach to move beyond the full-state method  and to emulate large quantum systems without requiring strong assumptions on the circuit and noise model. Originally developed by the condensed-matter physics community for studying strongly correlated systems, they have recently emerged as arguably the most successful way to emulate noiseless quantum circuits~\cite{vidal-tebd, PhysRevLett.127.040501, PhysRevLett.128.030501, PhysRevLett.133.230601, PhysRevA.96.062322, ayral2023, zhou2020limits}. The key benefit of tensor network methods is that the memory and runtime required to accurately emulate a quantum circuit scales only \textit{linearly} in the number of qubits, instead shifting the exponential scaling to the amount of entanglement built up by the circuit.\footnote{The scaling of the runtime of tensor network methods in the number of qubits can be somewhat worse than linear depending on the method used, but it is typically close to linear~\cite{fermioniq_ava}.} The fact that many quantum circuits do not maximally build up entanglement means that circuits acting on large numbers of qubits can often be accurately emulated in reasonable time~\cite{PRXQuantum.5.010308, PhysRevLett.129.090502}. 

These observations imply that tensor networks should, in theory, also be successful at emulating large-scale noisy quantum circuits, since noise makes quantum states less entangled, thus reducing computational requirements. Indeed, an early exploration of this topic in~\cite{noh_2020} showed that for noisy circuits entanglement builds up to a maximum -- the \emph{entanglement barrier} -- after which the noise slowly starts moving the state towards a low-entanglement fixed point (typically the maximally mixed state, if the noise is unital). All that is required to faithfully emulate a noisy circuit is to make it past the entanglement barrier without incurring significant approximation errors. 

However, crossing this barrier is made challenging by the fact that initially, alongside quantum entanglement, the noise itself builds up \textit{classical} correlations that also need to be captured by the tensor network state. The situation is further complicated by (i) the larger local physical dimension needed to represent a density matrix~\footnote{For a density matrix the individual sites correspond to a four-dimensional space, instead of simply being two-dimensional. The alternative is to use a trajectory-based approach. However, this introduces the overhead of sampling from many trajectories as well as the limitation that each individual trajectory is a pure-state emulation that (typically) continues to build up entanglement with increasing circuit depth, which is why in this work we stay within the vectorised density matrix framework.} and (ii) the fact that many tensor network algorithms (and more importantly their theoretical guarantees) do not directly carry over to emulation of mixed-states, and circumventing this seemingly requires the introduction of either additional computational overhead or uncontrolled errors into the simulation.\footnote{Uncontrolled errors arise due to loss of positive semi-definiteness (PSDness) of the density matrix. PSDness can be enforced in the tensor network using the locally purified form (LPF)~\cite{de2013purifications, werner2016positive} However, enforcing LPF, especially in the presence of gates that act on multiple tensors, introduces significant computational overhead, making the method much less performant than simply not enforcing PSDness but running at a higher bond dimension, which is why in this work we opt for the latter~\cite{de2013purifications, werner2016positive}. Consequently, to ensure that the state stays close to PSD, the simulation needs to be of high fidelity.}. For these reasons, tensor network emulations of noisy circuits -- especially in the low-noise regime where quantum and classical correlations are present simultaneously -- often struggle to achieve high fidelity. Unfortunately, this is precisely the most interesting regime: with particular relevance to error correction and mitigation.

Given a particular noisy simulation task with a challenging entanglement barrier to cross, we make the following observations:~1.~introducing extra noise lowers the barrier, and enables high-fidelity emulations;~2.~if the strength of the added noise is controlled by a single parameter, we find that expectation values of observables can be fitted to a simple function of that parameter. These two observations together imply that the inclusion of additional noise introduces a controllable error that can be approximately removed via extrapolation back to the low-noise regime of interest, thereby providing a method for accurate low-noise circuit emulations beyond full-state methods.

\noindent \paragraph*{Main contribution of this work} We propose and numerically validate a method for emulating noisy quantum circuits in the `difficult' regime: when the noise strength is small but non-zero. This regime is of particular relevance to quantum hardware and algorithm development on near-term devices, and for larger numbers ($\geq 40$) of qubits it is beyond the reach of full-state methods and typically very challenging for approximate (tensor network) emulators. 

Our method is based on the observation that noise destroys entanglement, combined with a modified version of a noise-mitigation technique from the quantum computing literature known as zero-noise extrapolation (ZNE). More precisely, our approach is to artificially add noise to the circuit to allow for high-fidelity emulations using tensor networks, and then extrapolate the results back to the point of interest. The key benefit of this technique, which we call `non-zero noise extrapolation' (NZNE), is that it allows us to avoid \textit{uncontrollable} approximations coming from low-fidelity tensor network emulations of mixed states by introducing instead a \textit{controllable} approximation in the form of extra noise, which then permits meaningful extrapolations to be made.\\

\subsection{Related work}

The literature on emulating quantum circuits using tensor network methods is by now very extensive, and we refer the reader to this survey paper~\cite{xu2023herculean} for a broad overview. On the other hand, the simulation of noisy quantum circuits using tensor networks is a comparatively under-explored topic. Noh et al.~\cite{noh_2020} consider the task of emulating 1D noisy random quantum circuits using matrix product operators (MPO). They do this primarily to study the entanglement entropy (with respect to the MPO, see Appendix~\ref{sec:mpo_entanglement_entropy}) of the state over time, and do not consider the extension of their method to simulating general quantum circuits in the presence of realistic noise. There is however a substantial literature addressed at the closely related problem of emulating open quantum systems, mainly from the condensed matter community \cite{verstraete2004matrix, PhysRevLett.93.207205, PhysRevResearch.5.033078, Jaschke_2018}. We note here that the standard way to represent a density matrix using an MPO has no computationally tractable way of enforcing or even checking that the resulting approximation satisfies the condition of positive semi-definiteness~\cite{kliesch2014matrix}. There are attempts to enforce this condition using the so-called locally-purified form, and we refer the reader to refs.~\cite{de2013purifications, werner2016positive} for detailed discussions on this method. Although it is an interesting Ansatz, and one that likely merits further attention, the locally-purified form is less expressive than a generic MPO, and leads to a significant increase in the computational cost of emulating a quantum system. For these reasons we will not use this approach or discuss it further in this work. 

As previously mentioned, our noise extrapolation approach is closely related to the technique of zero-noise extrapolation from the quantum computing literature \cite{temme2017error, li2017efficient}. The situation explored in this paper is simultaneously both simpler and more complicated than that considered in the case of real quantum hardware. On the one hand, in an emulation we have direct control over the strength of the noise (which makes it easy to take measurements at different noise strengths), and we do not experience statistical noise in the computation of observables since these can be computed exactly from a tensor network state. On the other hand, all the emulations that we can perform have errors coming from MPS-based compression of the vectorised density matrix: a source of errors that becomes very significant in the low noise regime. For the final extrapolation with respect to the noise strength we employ a simple exponential Ansatz, in line with other recent works~\cite{giurgica2020digital,endo2018practical,cai2021multi}.

Another related piece of work is ref.~\cite{rakovszky2020dissipationassistedoperatorevolutionmethod}. Here the authors make use of dissipation to reduce the required bond dimension needed to accurately simulate the MPO of an operator evolving in the Heisenberg picture, allowing them to compute spin and energy diffusion constants with high accuracy. Our approach is similar, but we make use of a different source of dissipation, inspired by quantum computing, and we also keep track of the entire density matrix of the evolving system in MPO form instead of a single observable.

\subsection{Organisation}
We begin by providing an explanation of our method in Section~\ref{sec:method}, which contains: (i) a high-level description of a tensor network method for emulation with density matrices (Section~\ref{sec:tn_emulator}); and (ii) an explanation of the non-zero noise extrapolation technique developed in this work (Section~\ref{sec:nzne}). Next, in Section~\ref{sec:results}, we apply our technique to a set of benchmark quantum circuits. We first compare to full state simulation on small systems (Section~\ref{sec:results_exact}), and then demonstrate how the method scales to large systems beyond the regime of exact simulation by full-state methods (Section~\ref{sec:large-systems}) .  In section~\ref{sec:discussion} we conclude with a short discussion and provide possible directions for future work.

\section{Method}\label{sec:method}

Tensor networks provide accurate and efficient representations of quantum states with limited entanglement between different subsystems. Noisy quantum computers are subject to decoherence, which bounds the extent to which entanglement can build up in the quantum state of the device. The fundamental idea behind our proposal is to make use of this effect to improve the accuracy of a tensor network emulation of a noisy quantum circuit. We do so by artificially introducing noise, which reduces the amount of entanglement built up by the circuit and, therefore, improves the accuracy of the emulation. The result of the emulation can then be used to compute local, few-body observables over a range of different noise strengths and extrapolate these into the regime of low noise. Here we validate this extrapolation method by comparing it to a direct tensor network emulation, the comparison was done with reference to exact methods (state-vector and matchgate simulations). Remarkably, we find that this extrapolation method performs well, and significantly outperforms a direct tensor network emulation in terms of accuracy, in the examples we have studied.

\subsection{Tensor network emulation for noisy quantum circuits}\label{sec:tn_emulator}
 In this section we fix notation for discussing tensor network based emulations of noisy quantum circuits, and comment on the heuristic fidelity that is generally obtained from such methods. 
\subsubsection{State representation}
We use a 1-dimensional tensor network Ansatz to approximate the density matrix of the state at all times during the computation, an approach also taken by~\cite{noh_2020}. More precisely, we consider an $n$-qubit density matrix $\rho$ in vectorised form:
\begin{align}
    \rho  = \sum_{i_1,..., i_n, j_1, ... j_n} \rho_{i_1,..., i_n, j_1,..., j_n}\ket{i_1,..., i_n}\bra{j_1,..., j_n}\\
    \Downarrow \hspace{5cm} \nonumber\\
    \kett{\rho} = \sum_{i_1,..., i_n, j_1, ... j_n} \rho_{(i_1, j_1)..., (i_n, j_n)}\kett{(i_1, j_1),..., (i_n, j_n)} \label{eqn:vectorized_dm}
\end{align}
where $i_k, j_k$ index the `ket' and `bra' indices for qubit $k$. In the above we have introduced the superket notation such that $\ket{i}\bra{j} \mapsto \kett{(i, j)}$.

We then use a matrix product state with local dimension 4 (storing both the ket and bra indices for each qubit) and bond dimension $D$ to represent the state:
\begin{align}
    \kett{\rho}_{(i_1, j_1),..., (i_n, j_n)} = \sum_{\alpha_1,...,\alpha_{n-1}=1}^D M^{[1]\, (i_1,j_1)}_{\alpha_1} M^{[2]\, (i_2,j_2)}_{\alpha_1, \alpha_2}...\,M^{[n]\, (i_n,j_n)}_{\alpha_{n-1}}\,
\end{align}
The tensors $M^{[1]}, M^{[2]}, \dots, M^{[n]}$ together constitute the tensor network Ansatz for the mixed state, which we term a `vectorised matrix product operator' (VMPO). Note that we do not refer to it as a `density' operator as we do not impose any additional constraints on the entries of the tensors, so the state encoded by it need not be positive semi-definite (PSD).\footnote{That the density matrix is PSD cannot even be checked efficiently given only the tensors $M^{[1]}, M^{[2]}, \dots, M^{[n]}$~\cite{kliesch2014matrix}.}

\subsubsection{Fidelity}\label{ssec:fidelity}
Tensor network emulations are approximate emulations. The degree of approximation is determined by the \emph{bond dimension} of the VMPO. Each emulation comes with a notion of accuracy, given by its \emph{fidelity}.

\paragraph{Setup}
Given a noisy circuit $C$ consisting of a sequence of $T$ completely positive trace preserving (CPTP) operators $C_1,\dots,C_T$, individually referred to as \emph{subcircuits}, and an initial state $\rho_0$, let $\rho_l = C_l \cdots C_1(\rho_0)$ denote the state of the circuit after having applied the first $l$ subcircuits to $\rho_0$. The goal of emulating the circuit $C$ on initial state $\rho_0$ is to obtain an approximation of $\rho_T$.

\paragraph{Notation}
With a slight abuse of notation, we will use $\kett{\rho_l} = C_l \cdots C_1 \kett{\rho_0}$ to denote the state $\rho_l$ in its vectorised representation. 
Furthermore, given two mixed states $\sigma, \sigma'$, let $\bbrakett{\sigma}{\sigma'}$ denote the usual inner product between the vectors $\kett{\sigma}$ and $\kett{\sigma'}$. Note that this is equivalent to the Frobenius inner product $\langle \sigma, \sigma' \rangle_F = \Tr(\sigma^\dagger \sigma')$ of the two density matrices $\sigma, \sigma'$. Finally, note that $\bbrakett{C_{l+1}(\rho_l)} {C_{l+1}(\rho_l)} \neq \bbrakett{\rho_l}{\rho_l}$ in general, since CPTP operators acting on density matrices need not preserve their Frobenius norm.

\paragraph{Approximate emulation}
Using any tensor network compression method (e.g.~\cite{vidal-tebd, ayral2023}), for every $l$ we obtain $\kett{\tilde{\rho}_l}$, the approximation to $\kett{\rho_l}$, by finding the closest (in Frobenius distance) fixed-bond-dimension $D$ state with Frobenius norm 1 to the the state $C_l\kett{\tilde{\rho}_{l-1}}$; the latter of which is obtained by applying $C_l$ to the \textit{approximate} state $\kett{\tilde{\rho}_{l-1}}$. The initial state is always a product state so we always start with $\kett{\tilde{\rho}_0} = \kett{\rho_0}$. We introduce the \emph{partial fidelity} $f_l$ for the $l$-th subcircuit, defined by
\begin{equation}
    f_l := \frac{|\bbrakett{\tilde{\rho}_{l}}{C_l|\tilde{\rho}_{l-1}}|^2}{||\tilde{\rho}_{l}||^2_2 \, ||C_l\tilde{\rho}_{l-1}||^2_2}\,
\end{equation}
where $||.||_2$ is the Frobenius norm (or 2-norm). The partial fidelity, $f_l$, expresses how well the emulation managed to approximate the application of the $l$-th subcircuit on the state $\kett{\tilde{\rho}_{l-1}}$. The partial fidelity $f_l$ lies in the range $[0, 1]$, and when it is equal to 1 the $l$-th subcircuit was applied exactly to $\kett{\tilde{\rho}_{l-1}}$.\footnote{Up to normalization, since the state $\kett{\tilde{\rho}_l}$ will have Frobenius norm 1 instead of trace norm 1.}

\paragraph{Emulation fidelity}
The quantity that describes how well the emulation performed as a whole  is the \textit{true fidelity} $F$ of the emulation, defined by

\begin{equation}
    F := \frac{|\bbrakett{\tilde{\rho}_{T}}{\rho_{T}}|^2}{||\tilde{\rho}_{T}||^2_2 \, ||\rho_{T}||^2_2}\,.
    \label{eq:true_fidelity}
\end{equation}

We cannot compute the true fidelity exactly, but we can estimate it with the \textit{emulation fidelity} $\mF$, which we take to be the product of all partial fidelities:
\begin{equation}
        \mF := \prod_{l=1}^T f_l\,.
        \label{eq:emulation_fidelity}
\end{equation}

$\mF$ is generally believed to be a good approximation to $F$, though this is not guaranteed.\footnote{Although note that in the special case where $\mathcal{F} = 1$ we are guaranteed that $F=1$, since every subcircuit was emulated exactly. } There are heuristic arguments that corroborate this so long as the errors introduced by the approximations $\tilde{\rho}_l$ of $\rho_l$ are not overly `structured', and this is also backed up by significant numerical evidence~\cite{zhou2020limits}. Indeed we have found $\mF$ to be a good approximation of $F$ for the circuits we consider in this paper within the regime accessible to exact simulations (see e.g. Appendix~\ref{sec:fidelity_true_fidelity}).

Finally, it is also worth noting that even if the true fidelity $F$ is high, we are not guaranteed that the expectation values of observables computed from the states $\rho_T$ and $\tilde{\rho}_T$ are close. It is possible to construct observables whose expectation values in the two density matrices differ proportionally to the trace distance$||\rho_T -\tilde{\rho}_T||_1$, which can be bounded only very loosely by the Frobenius distance. However, observables that differ proportionally to the trace distance are highly atypical. In fact, it has been argued in ref.~\cite{Clark_dm} that the difference between expectation values of typical observables in the two states $\rho_T$ and $\tilde{\rho}_T$ is proportional to the Frobenius distance $||\rho_T - \tilde{\rho}_T||_2$.\footnote{The argument provided in ref.~\cite{Clark_dm} was presented in the context of using tensor network methods for approximation of a classical probability distribution (instead of a density matrix), however, the same issue arises in that context, and their argument is also relevant here.} Given this, we can expect that the emulation fidelity (which is closely related to the Frobenius distance) should provide a reasonable guide to the size of the error of the expectation values of typical observables, even if it is a poor guide to the difference in expectation values in the atypical worst-case scenario. We have found that, in practice, high emulation fidelity means that most observables can be approximated remarkably well.

\subsection{Non-zero noise extrapolation}\label{sec:nzne}

The non-zero noise extrapolation technique makes use of tensor network emulations of a noisy quantum circuit, acting on a fixed initial state. The noise model of the quantum circuit is assumed to be tuneable, with a parameter $\lambda$ that controls the strength of the noise. For a range of values of $\lambda$, tensor network emulations are used to provide an estimate of the expectation value of an observable $O$ evaluated in the state generated by applying the noisy circuit at noise strength $\lambda$ on the given initial state. We denote the expectation value which is being estimated by $\langle O\rangle_\lambda$. We make the following crucial assumption about the accuracy of the tensor network estimates that we obtain from these emulations:

\begin{itemize}
    \item[] \textbf{Accuracy} The accuracy of the emulation (in terms of $\mF$) should increase with increasing noise strength $\lambda$. 
\end{itemize} 
\begin{figure*}[t!]
\centering
\includegraphics[width=.5\textwidth]{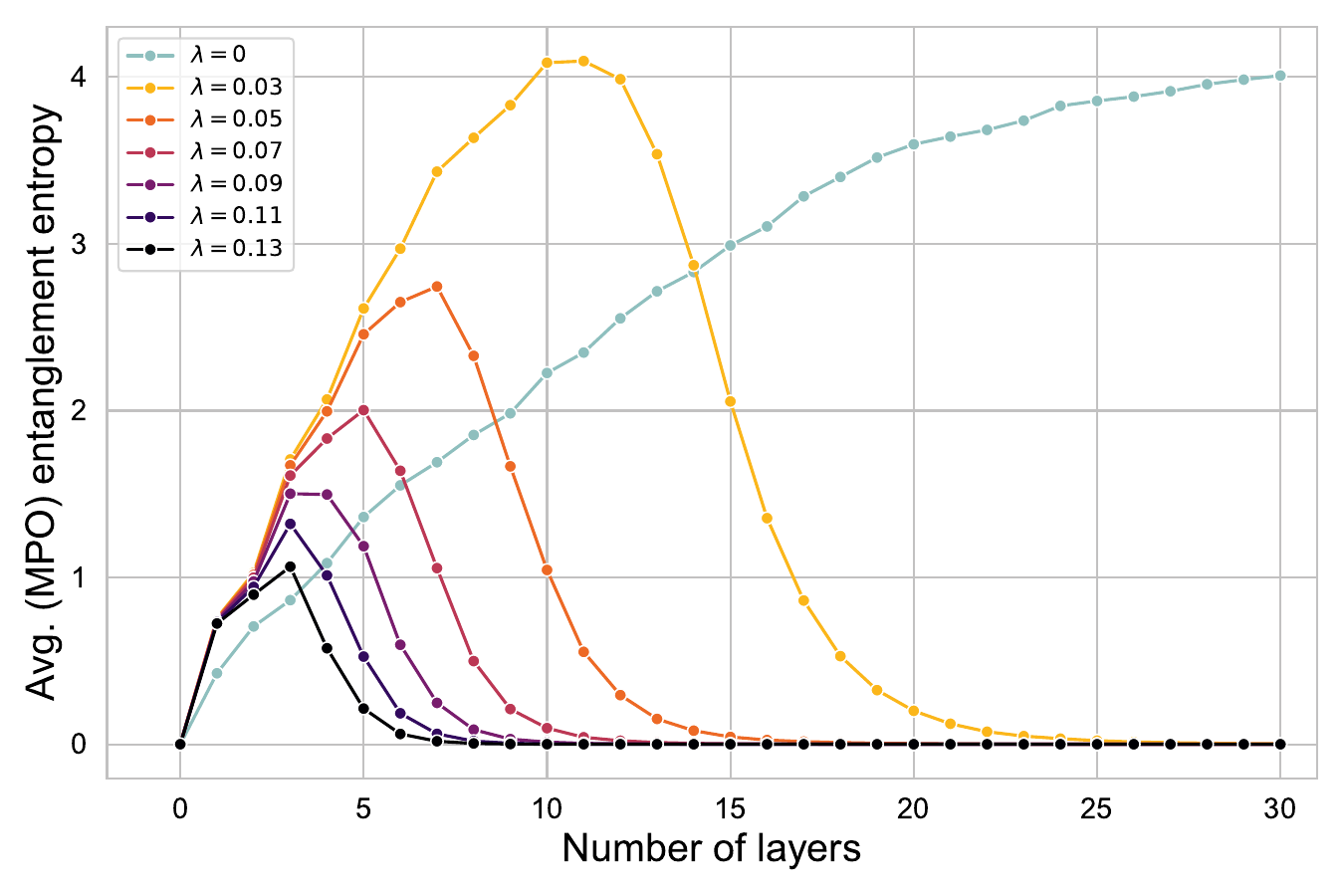}\hfill
\includegraphics[width=.5\textwidth]{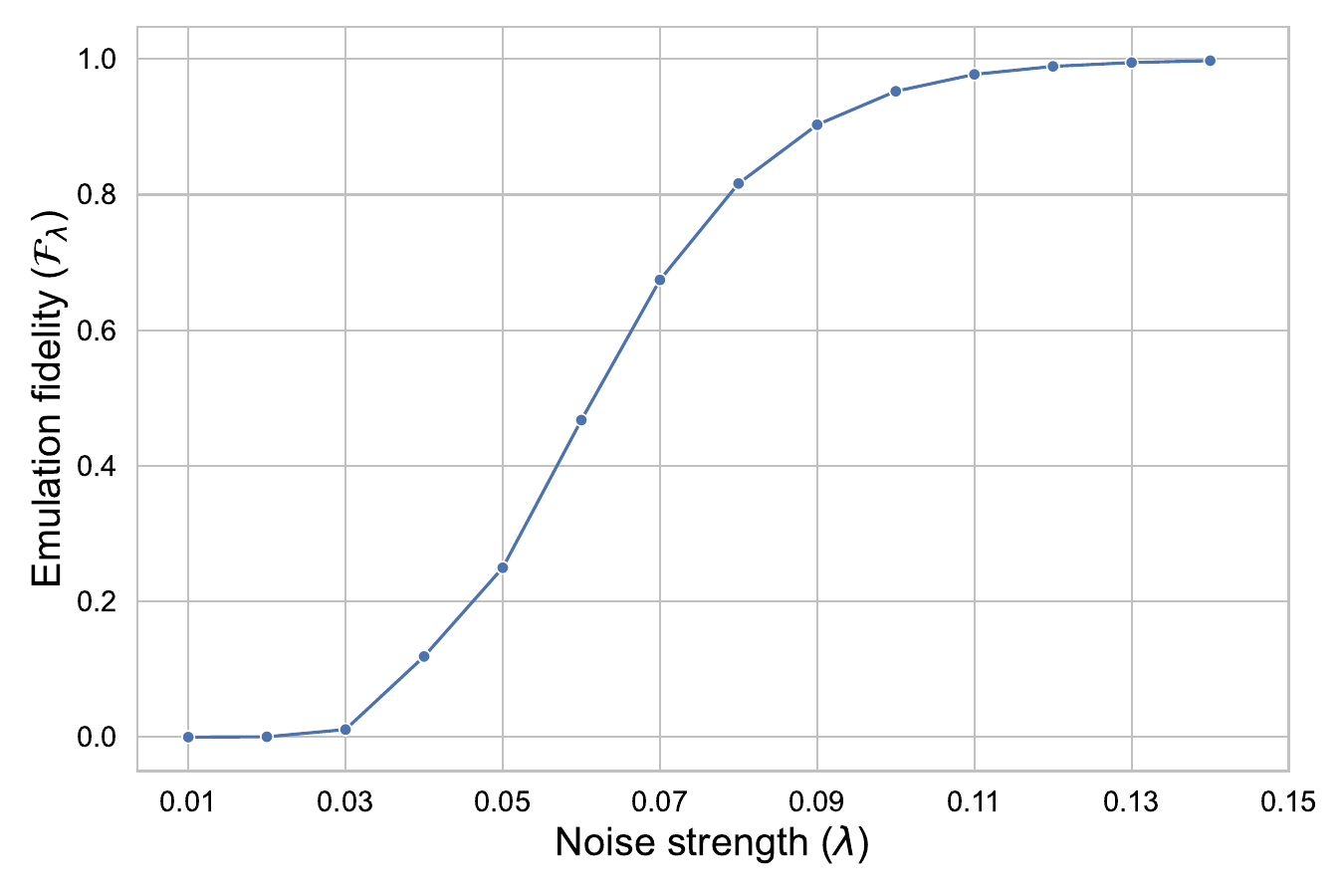}\hfill
\caption{\textbf{Left}: Average (over 10 runs) of the entropy of the quantum state after each layer random quantum circuits on 16 qubits. Each layer of the circuit consists of Haar-random single-qubit gates followed by controlled-$Z$ gates on all neighbouring pairs of qubits with 1D connectivity. For $\lambda > 0$, the entanglement entropy is the MPO entanglement entropy of the VMPO representing the state. For $\lambda = 0$ the entanglement entropy is the `usual' entanglement entropy of the pure state (represented by a matrix product state). For both, the entropy is measured at a cut across the middle 2 qubits.
\textbf{Right}: Emulation fidelity $\mF_\lambda$ for various noise strengths $\lambda$ of the depolarizing noise model whist keeping the bond dimension fixed (to 1500). The circuit is a random 2D quantum circuit on 25 qubits, consisting of 30 layers of: Haar-random single-qubit gates on all qubits followed by controlled-$Z$ gates on all neighbouring pairs of qubits. All circuits were applied to the all-zeros initial (pure) state.}
\label{fig:accuracy_v_lambda}
\end{figure*}

\noindent As mentioned in the introduction, the observation that the accuracy increases with increasing noise strength is the key observation that allows our technique to work. That tensor network emulations behave like this has long been `folklore', however, we particularly draw attention to ref.~\cite{noh_2020}, which studies the MPO-entanglement entropy (Appendix~\ref{sec:mpo_entanglement_entropy}) of a density matrix as it evolves under noisy quantum circuits. The authors of \cite{noh_2020} conclude that in noisy quantum circuits this quantity is bounded, with a peak that decreases as the strength of noise increases. Moreover, low MPO entanglement entropy (across all possible bi-partitions of the system that respect the MPO qubit-ordering) implies that it is possible to accurately represent the density matrix with modest bond dimension. Taken together, these observations imply that when the strength of noise is higher we expect to see a significant improvement in the accuracy of a tensor network emulation. 

In Fig.~\ref{fig:accuracy_v_lambda} we illustrate this behaviour for random circuits with a noise model consisting of depolarizing channels. We verify both that the MPO entanglement entropy of the state is bounded with a peak that decreases for higher levels of noise (Fig.~\ref{fig:accuracy_v_lambda}: left) and that the (heuristic) emulation fidelity improves as the noise strength increases (Fig.~\ref{fig:accuracy_v_lambda}: right).

\subsubsection{Parametrised noise models}
For simplicity in defining and working with the extrapolation method we are proposing, in this work we consider noisy quantum circuits which have a well-defined parametrised noise model associated with them. These are the simplest cases on which this method could be applied, but they are not the only cases. We leave it to future work to investigate how the proposed method can be extended to other types of noise model (such as by artificially adding tuneable noise to a non-tuneable noise model).

In particular, we always consider noise models built from CPTP channels (noise channels) on one or two qubits, which we add to the circuit by post-composing each gate of the circuit with the corresponding channel:
\begin{equation}
    U \to C \circ U
\end{equation}
where $U$ is some unitary channel representing the noiseless gate, and $C$ is the noise channel. Moreover, we let every channel in the noise model depend on some parameter $0 \leq \lambda \leq 1$ which quantitatively controls the gate fidelity and decoherence when that channel acts on typical gates and quantum states, with $\lambda = 0$ corresponding to the case when there is no noise in the circuit. 

There are many noise models that behave in this way, including elementary examples such as one- and two-qubit depolarizing channels, and single-qubit dephasing channels, as well as more complex noise models such as the noise model for cat-qubits provided in refs.~\cite{mirrahimi2016cat, guillaud_2021, guillaud2023quantum}.

\subsubsection{Implementation}\label{ssec:implementation}
Our method involves taking a quantum circuit $C$ with a family of noise models $\mathcal{N}(\lambda)$ parametrised by $\lambda$. Given an observable $O$ of interest, for any noise strength $\lambda$, we denote the expectation value of the state obtained by applying the circuit subject to noise model $\mathcal{N}(\lambda)$ to the given initial state by $\langle O\rangle_\lambda$. 

Our goal is to estimate  $\langle O\rangle_{\lambda^*}$ at some given target noise strength $\lambda^*$ that cannot be emulated with high fidelity. To start with, we run emulations for a range of values of $\lambda$, including some emulations that have high fidelity (and are likely to be accurate) and others which have lower fidelity: we will denote the set of included values of $\lambda$ by $\Lambda$. We then perform the following two steps. First, because our emulations are approximate emulations, we (A) extrapolate fidelity to one to obtain the extrapolated expectation values $\overline{\langle O \rangle}_{\lambda}$ for every $\lambda \in \Lambda$. In this step we also remove all values of $\lambda$ from $\Lambda$ for which the extrapolation is not reliable. Second (B), using the obtained fidelity-extrapolated data points $\{\overline{\langle O \rangle}_{\lambda}\}$ we extrapolate in noise strength to the target noise strength $\lambda^*$ (which may itself be included in $\Lambda$, but need not be) to obtain our desired estimate of $\langle O\rangle_{\lambda^*}$. 

\paragraph*{(A) Fidelity extrapolation}\label{para:fidelity_extrapolation}
Given $\lambda \in \Lambda$, we compute expectation values for a range of different bond dimensions $D_1, ..., D_k$. For every $j \in \{1, \ldots, k\}$ we will denote the expectation value computed using bond dimension $D_j$ by $\langle O \rangle_\lambda^{(D_j)}$, and the corresponding emulation fidelity by $\mathcal{F}_\lambda^{(D_j)}$. For clarity, we will also refer to the maximum bond dimension $D_k$ used as $D_{\max}$, and the corresponding expectation value and fidelity at noise strength $\lambda$ as $\langle O \rangle_\lambda^{\max}$ and $\mathcal{F}_{\lambda}^{\max}$ respectively.  

Next, we perform an extrapolation in fidelity (to one)~\footnote{Extrapolating fidelity to one corresponds to extrapolating bond dimension to infinity -- a limit that is often taken in the tensor network literature. We prefer to extrapolate in fidelity because we find it gives more accurate results. The reason for this is that fidelity is more informative: it provides information on the accuracy of the emulation, whereas bond dimension merely signifies the amount of computational resources used.} via a straight-line fit to obtain an extrapolated value $\overline{\langle O \rangle}_\lambda$, which is our approximation of $\langle O \rangle_{\lambda}$. Relying on extrapolated values introduces the possibility of numerical errors if the observable of interest converges poorly. To avoid this, we introduce the following restriction on the set $\Lambda$:
 \begin{enumerate}
     \item \label{crit:conv} We only include in $\Lambda$ those values of $\lambda$ for which the values $\langle O \rangle_\lambda^{(D_j)}$ have approximately converged by the time $D_j$ has reached $D_\mathrm{max}$.
\end{enumerate}
What counts as ``approximately converged'' could be determined in many different ways. However, in practice we have found it useful to make use of a quantitative measure which is explained in more detail in Appendix \ref{sec:criteria-convergence}. We note that, because we can use higher bond dimension for the emulation of the noiseless circuit than we can use for the noisy emulations, in practice the pure state point $\overline{\langle O \rangle}_{\lambda = 0}$ is typically always approximately converged.

\paragraph*{(B) Noise strength extrapolation}
Following refs. \cite{cai2021multi, endo2018practical, giurgica2020digital} we achieve the second extrapolation by fitting the data points to an exponential Ansatz:
\begin{equation}
    \hat{O}(\lambda) = a e^{- b\lambda} +c \, .
\label{eqn:exp_Ansatz}
\end{equation}
 
There are many ways to fit an exponential Ansatz such as that in equation~\eqref{eqn:exp_Ansatz}, but by far the two most common are: (1) simply fitting the parameters $a, b, c$ directly by optimizing some choice of loss function $\mathcal{L}$, or (2) if $c=0$ it is possible to fit the log of the absolute values of the data points, $\rm log (|\langle \overline{O}\rangle_\lambda|)$, to a straight line (a log-linear fit). This straight line can be fitted by optimizing a suitable choice of loss function $\mathcal{L}$, which is usually taken to be the mean-square error (but need not be).

In this paper we make use of both approaches. We primarily rely on the second approach (a log-linear fit), making use of a tailored loss function. The loss function that we use is a weighted mean-square error, where the weights are chosen to emphasise points $\langle \overline{O}\rangle_\lambda$ where $\lambda$ is close to $\lambda^*$ and also those where `$\mF_{\lambda}^{\max} \gg 0$'. More details on the precise form of loss function we use can be found in Appendix~\ref{ssec:loss-function-details}.  However, to apply this method it is necessary to introduce two further restrictions to the set of points $\Lambda$ being used in the fit:

\begin{enumerate}
\setcounter{enumi}{1}
     \item \label{crit:small} For numerical stability, we do not include any values of $\lambda$ for which
     $|\overline{\langle O \rangle}_{\lambda}|$ is too small (in this work we require that $|\overline{\langle O \rangle}_{\lambda}|\geq 10^{-10}$).
     \item \label{crit:sign} We do not include any values of $\lambda$ for which $\overline{\langle O \rangle}_{\lambda}$ are of the opposite sign of the value computed at zero noise strength $\overline{\langle O \rangle}_{\lambda = 0}$, to ensure the valid applicability of the exponential Ansatz~\eqref{eqn:exp_Ansatz} with $c = 0$.  
\end{enumerate}

 Together with criterion~\ref{crit:conv} above, these will be referred to as criteria~\ref{crit:conv}-\ref{crit:sign}. In situations where these restrictions lead to an insufficient number of values in $\Lambda$ to perform a meaningful extrapolation, we will instead fall back on the first approach (a direct exponential fit, fitting all three parameters $a, b, c$ in  Ansatz~\eqref{eqn:exp_Ansatz}), using the mean-square error as our loss function (dropping the criteria~\ref{crit:small}-\ref{crit:sign} while keeping criterion~\ref{crit:conv})\footnote{The choice to rely primarily on the log-linear fit over the exponential fit was not guided by a principled justification, but simply by the practical consideration that we found it more straightforward to design an appropriate tailored loss function for the log-linear fit. We leave it as a project for future work to perform a more in depth analysis of what the optimal approach should be.}. However, there are some (rare) occasions where it is not possible to perform an extrapolation in the noise strength using either method, due to insufficient number of approximately converged, fidelity extrapolated data-points (i.e. points passing through criterion~\ref{crit:conv}). In these cases, it is not possible to apply the non-zero noise extrapolation method, and so we fall back on simply performing a tensor network emulation at the target noise strength.

\begin{figure*}[t!]
    \centering
    \subfigure{\label{fig:a}\includegraphics[width=0.5\textwidth]{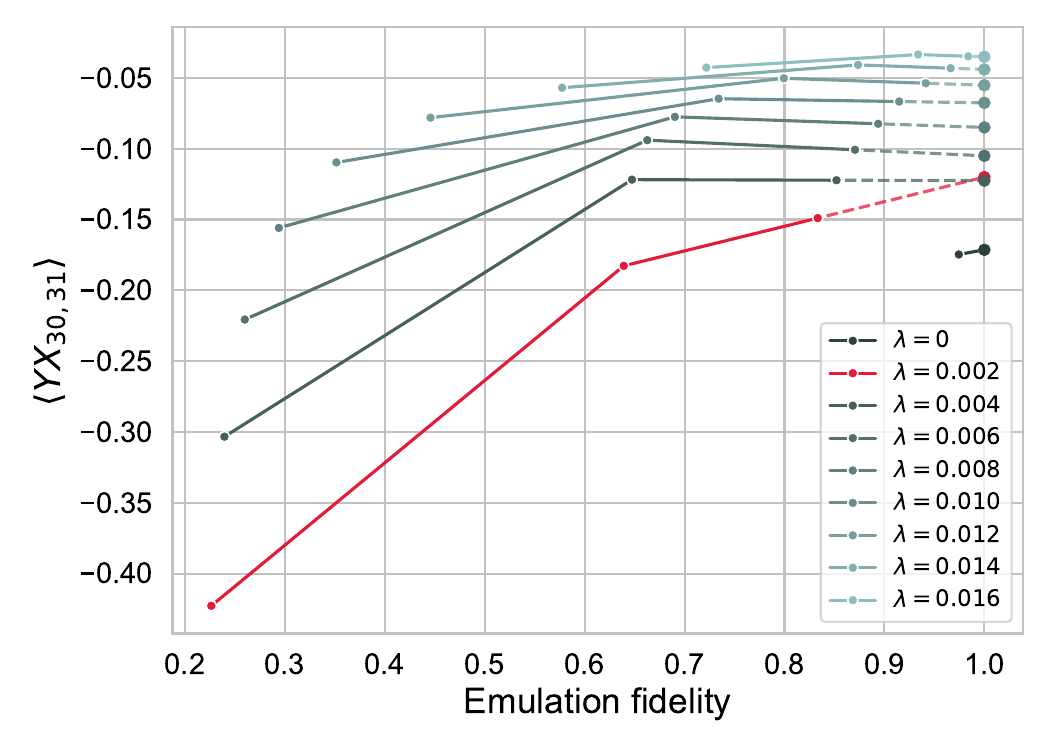}}%
    ~
    \subfigure{\label{fig:b}\includegraphics[width=0.5\textwidth]{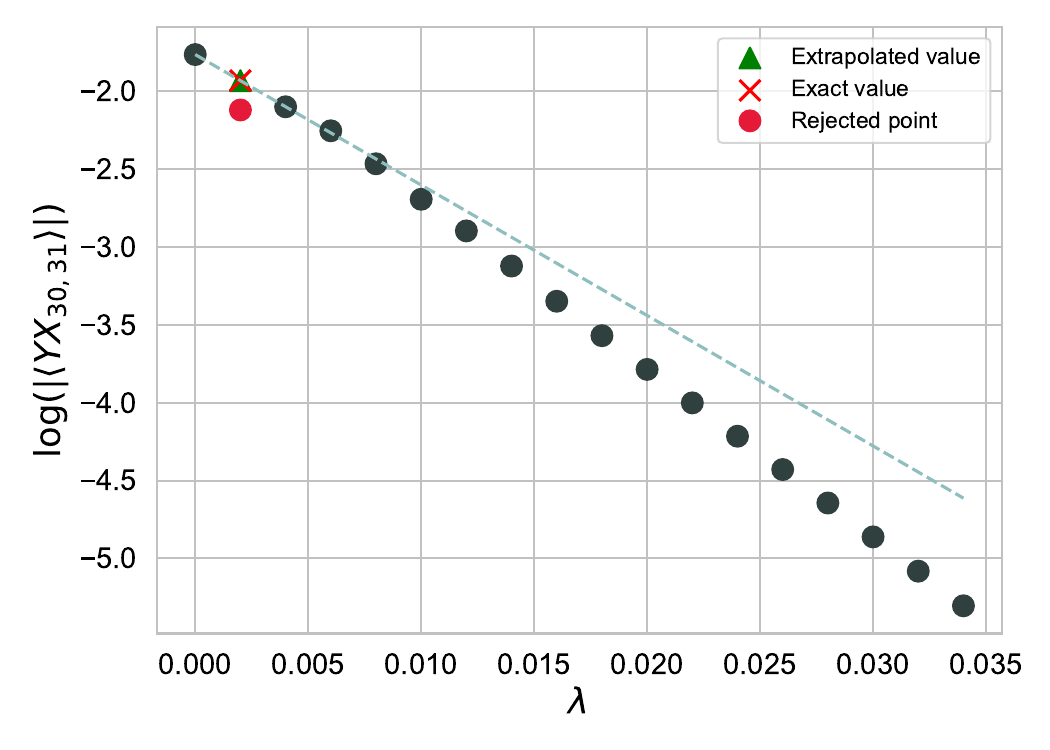}}

    \caption{Demonstration of the method for the observable $YX_{30,31}$ on a $60$-qubit instance of the $XY$ model (see main text or Section~\ref{app:xym} for details of the circuit). First we perform fidelity extrapolation to obtain estimates of $\overline{\ex{YX}}_{30, 31}$ at different noise strengths $\lambda$. For each $\lambda$ the fidelity extrapolation is obtained from 4 tensor network emulations with different bond dimensions $D \leq 300$. (a). Any poorly extrapolated results are removed ((a) red line). The remaining points are used for non-zero noise extrapolation, which consists of a (weighted) straight line fit applied to the logs of the absolute values of the expectation values (b). The weighted fit puts more emphasis on high-fidelity data points close to the target noise strength. The method is most useful when the emulation fidelity at the target noise strength is low, which was the case here.} 
    \label{fig:extrapolation_demo}
\end{figure*}

\subsubsection{Example}
We now give an example application of our technique. 
The circuit that we will use implements trotterised time evolution of the 1D $XY$ model on $60$ qubits, with Trotter step size $dt = 0.1$ and 30 Trotter steps. The initial state was chosen to be an anti-ferromagnetically aligned 1-D product state $\ket{1010....}$. Further details on the Hamiltonian and precise values of all parameters used can be found in Appendix~\ref{app:xym}. This circuit was chosen since it can be classically simulated to obtain exact results, despite being out of reach of full-state methods (see Section~\ref{ssec:xym_results} and Appendix~\ref{app:xym} for more on this point). The noise model is a stochastic matchgate noise model $\nmg$, in which single-qubit gates are noiseless and two-qubit gates are each followed by a two-qubit channel with strength $\lambda$ that applies all matchgate Paulis with equal probability. For a more detailed explanation of this noise model see Appendix~\ref{app:matchgate-noise}. For this example we take the target noise strength to be $\lambda^*=0.002$, and we take the observable whose expectation value we will estimate at this noise strength to be $O = YX_{30,31}$. 

To more clearly demonstrate the steps of our method, we artificially restrict the bond dimension to relatively low values: $D \leq 300$ , resulting in emulation fidelity of $\mF_{0.002}^{D=300} = 0.8332$. For each noise strength $\lambda$, we perform fidelity extrapolation as in Section~\ref{ssec:implementation} to obtain more accurate estimates of the expectation value at that noise strength (Figure~\ref{fig:extrapolation_demo} (a)). We select which data points to keep according to the criteria~\ref{crit:conv}-\ref{crit:sign}. For this example, applying the selection criteria led to the rejection of a single data point at $\lambda=0.002$ (Figure~\ref{fig:extrapolation_demo}~(a), in red), which converged poorly, and indeed one can see in Figure~\ref{fig:extrapolation_demo}~(b) that the data point of interest (red circle) is indeed an anomalous point. From the remaining data points, we perform a weighted straight line fit to their log values, and read off the extrapolated expectation value at the target noise strength (green triangle in the figure). This can be compared to the near-exact values (red cross) as computed by a trajectory simulation, from which we can conclude that the extrapolated value in Figure~\ref{fig:extrapolation_demo}~(b) is indeed much more accurate than the data point at $\lambda=0.002$ (red dot). Precisely, performing NZNE improved the relative error compared to a bond-dimension extrapolation from $0.1724$ to $0.0006$  for estimating $YX_{30,31}$.

\section{Results}\label{sec:results}
In this section we demonstrate our method for a variety of benchmark circuits and noise models, both for small and large systems, with varying qubit layouts (square or elongated) and boundary conditions (open or periodic). For the small circuits, we compare our results to exact simulations in order to study the accuracy of our method. For the large circuits, we investigate scaling behaviour and, for one particular benchmark circuit, compare again to exact results to provide evidence that our method continues to work well for large systems. For all emulations, we use Fermioniq's tensor network circuit emulator Ava~\cite{fermioniq_ava}, which allowed us to accurately emulate circuits on large numbers of qubits on a single NVIDIA Grace-Hopper GPU. 

We use three different circuits for benchmarking inspired by near-term algorithms for studying physical systems, all focused around the challenging task of time-evolving quantum systems, something that is natural for quantum computers but challenging for quantum emulators. 
\begin{itemize}
    \item \textit{Ising model (TFIM)}: Trotterised time evolution of the 2D transverse-field Ising model (TFIM) with periodic boundary conditions. Details of the model and circuit that we used are in Appendix~\ref{app:tfim}.
    \item \textit{Fermi-Hubbard model (FHM)}: Trotterised time evolution of the 2D Fermi-Hubbard model with open boundary conditions. Details of the model and circuit that we used are in Appendix~\ref{app:fhm}.
    \item \textit{$XY$ model (XYM)}: Trotterised time evolution of the (isotropic) $XY$ model in one dimension. This model is exactly solvable, and circuits implementing its time evolution map to so-called matchgate circuits~\cite{brod2016efficient,knill2001fermionic,jozsa2008matchgates}, which allows us to simulate the circuits classically for large numbers of qubits, including noise. Details of the model and circuit that we used are in Appendix~\ref{app:xym}.
\end{itemize}
We also use three different noise models for our benchmarks:
\begin{itemize}
    \item \textit{Depolarizing noise model}: A simple model of depolarizing noise applied after every two-qubit gate. See Appendix~\ref{app:depolarizing} for the definition.
    \item \textit{Cat-qubit noise model}: A noise model originating from the kind of biased noise present in quantum computers using Cat-qubits~\cite{mirrahimi2016cat,guillaud_2021,guillaud2023quantum}\,. See Appendix~\ref{app:cat_noise} for the definition. 
    \item \textit{Matchgate depolarizing noise}: A modified version of the depolarizing noise model for matchgate circuits, where matchgate Pauli operators are applied with equal probability following a two-qubit gate. See Appendix~\ref{app:matchgate-noise}.
\end{itemize}
Details about the observables whose expectation values we wish to estimate will be given in the relevant sub-sections below.

\subsection{Comparison to exact results}\label{sec:results_exact}

We first benchmark non-zero noise extrapolation on systems of up to $16$ qubits, so that we can compare the results to exact (full density matrix) simulations. We artificially lower the fidelity of the tensor network emulations by using small bond dimensions, which allows us to use the results obtained to give an indication of the performance on larger systems when similar emulations fidelities would be achieved.

For these comparisons, we consider a 14-qubit TFIM instance under depolarizing and cat-qubit noise, and a 16-qubit FHM instance under depolarizing noise. For each system we choose (by hand) a target noise strength $\lambda^*$ and maximum bond dimension $D_{\max}$ so that the corresponding fidelity satisfies $\mF_{\lambda^*}^{\max} \ll 1$ whilst $\mF_\lambda^{\max} \geq 0.99$ for $\lambda \gg \lambda^*$, allowing room for our extrapolation method to yield increased accuracy for the $\lambda^*$ data point over the one obtained by the actual emulation(s) performed at $\lambda^*$ itself. To quantify the improvement in accuracy, we compare the value obtained via non-zero noise extrapolation to the value from exact simulation as well as with the expectation value obtained from a single $D_{\max}$ emulation at $\lambda^*$. Note that we do not compare against the fidelity extrapolated expectation values at $\lambda^*$, because we have found that, at $\lambda^*$, fidelity extrapolation tends to give poorer accuracy than simply taking the expectation value obtained by the highest fidelity emulation (due to the fact that the emulation fidelity $\mF_{\lambda^*}$ is usually quite low and doesn't yield converged data points in the extrapolation).

\subsubsection{Transverse field Ising model (TFIM)}
We consider a 2x7 instance of the TFIM with periodic boundary conditions, and compute all nearest-neighbour $ZZ$-expectation values after 10 (second-order) Trotter steps of size $dt = 0.25$, starting from the all zero initial state (see Appendix~\ref{app:tfim} for more details). We set the maximum bond dimension $D_{\max}=32$ for all emulations (including pure-state). The qubits in the 2x7 lattice are column-major ordered (i.e.~the first column contains qubits 0 to 6, and the second 7 to 13).

\begin{figure*}
    \begin{minipage}[b]{\columnwidth}
    \centering
    \includegraphics[width=0.8\textwidth]{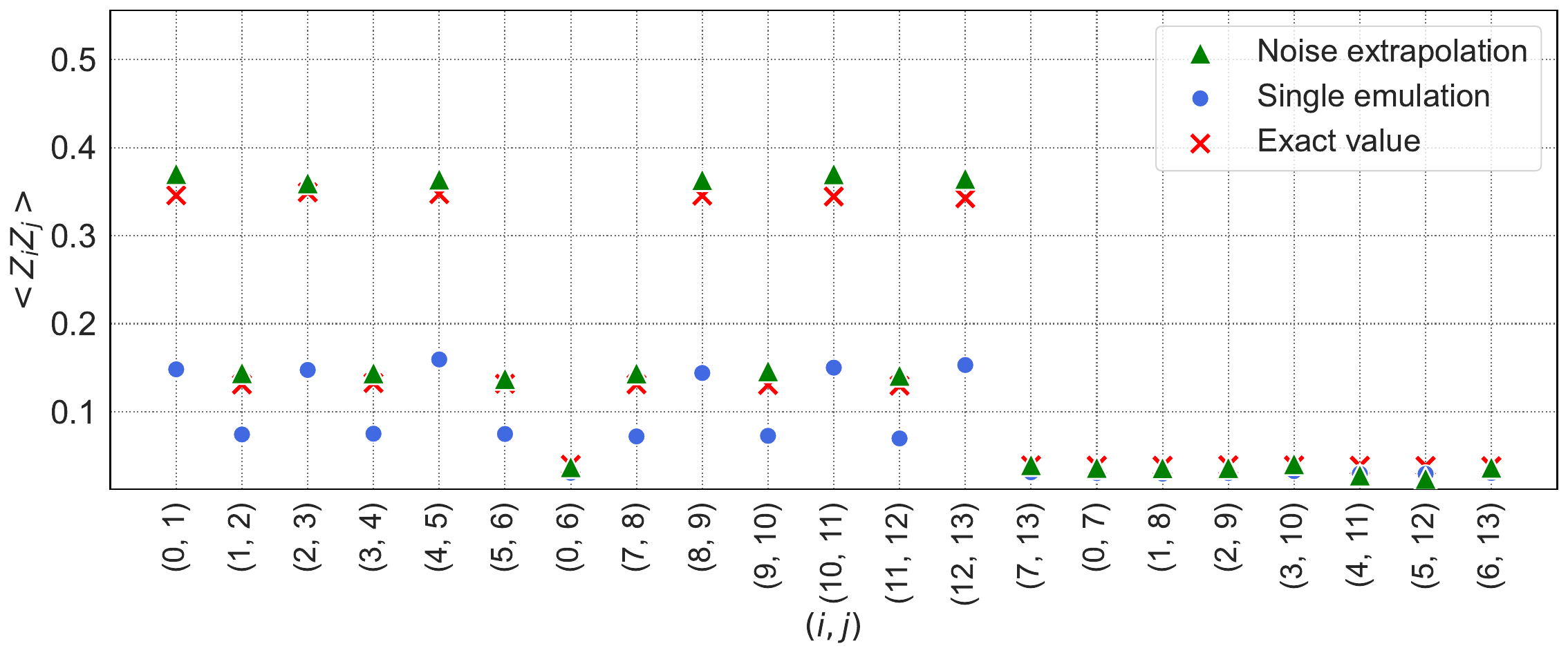}
    \caption{$\ex{Z_iZ_j}$ after $10$ Trotter steps of a $2 \times 7$ instance of the TFIM with depolarising noise at $\lambda^*=0.01$. A single emulation with $D_{\max}=32$ (and $\mathcal{F}_{\lambda^*}^{\max}=0.39$) yields an average absolute error of $0.0590$ compared to the exact values. Non-zero noise extrapolation with $D\leq32$ yields an average absolute error of $0.0093$.}
    \label{fig:dep_ising_14_scatter}
    \end{minipage}
    \hfill
    \begin{minipage}[b]{\columnwidth}
    \centering
    \includegraphics[width=0.8\textwidth]{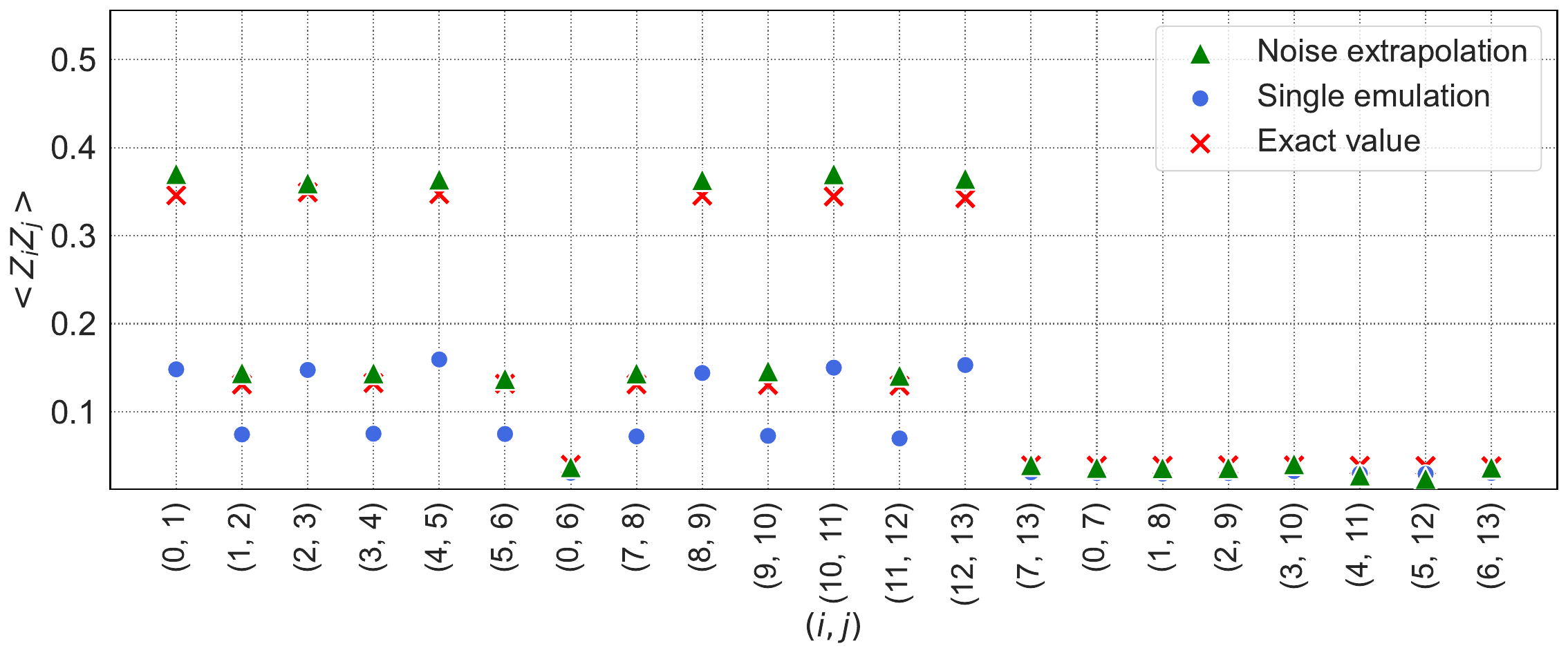}
    \caption{$\ex{Z_iZ_j}$ after $10$ Trotter steps of a $2 \times 7$ instance of the TFIM with cat noise at $\lambda^*=0.0005$. A single emulation with $D_{\max}=32$ (and $\mathcal{F}_{\lambda^*}^{\max}=0.29$ ) yields an average absolute error of $0.0696$ compared to the exact values. Non-zero noise extrapolation with $D\leq32$ yields an average absolute error of $0.0085$.}
    \label{fig:cat_ising_14_scatter}
    \end{minipage}
\end{figure*}

For the target depolarizing noise strength, which we set to $\lambda^* = 0.01$, we obtain $\mF_{\lambda^*}^{\max} = 0.39$, and for larger noise strengths $\lambda \geq 0.11$ we have $\mF_\lambda^{\max} \geq 0.99$. Figure~\ref{fig:dep_ising_14_scatter} shows the values for $\ex{ZZ}_{\lambda^*}$ on all neighbouring qubit pairs for depolarizing noise with $\lambda^* = 0.01$. We find that non-zero noise extrapolation obtains values of $\ex{ZZ}$ that are on average 6 times closer to the exact values than those obtained via a single $D_{\max}$ emulation at $\lambda^*$. 

Similar results are obtained in the case of cat noise with $\lambda^* = 0.0005$ as shown in Figure~\ref{fig:cat_ising_14_scatter} -- here non-zero noise extrapolation yields values that are over 8 times closer than those from a single emulation. 

To investigate how the accuracy of non-zero noise extrapolation behaves as a function of emulation fidelity, we consider the observable $\bar{E}_{\text{TFIM}}$ corresponding to the energy per site (which contains all the neighboring ZZ-terms, and is easier to plot than all ZZ-terms separately):
\begin{equation}
    \bar{E}_{\text{TFIM}} := \frac{1}{n} H_{\text{TFIM}}
\end{equation}
where $n$ is the number of qubits and $H_{\text{TFIM}}$ is the TFIM Hamiltonian (see Appendix~\ref{app:tfim} for the definition). To obtain an estimate for $\ex{\bar{E}_{\text{TFIM}}}$, we obtain estimates individually for each $\ex{ZZ}$ and $\ex{X}$ via non-zero noise extrapolation. 

\begin{figure*}[t!]
    \centering
    \subfigure{\label{fig:a}\includegraphics[width=0.5\textwidth]{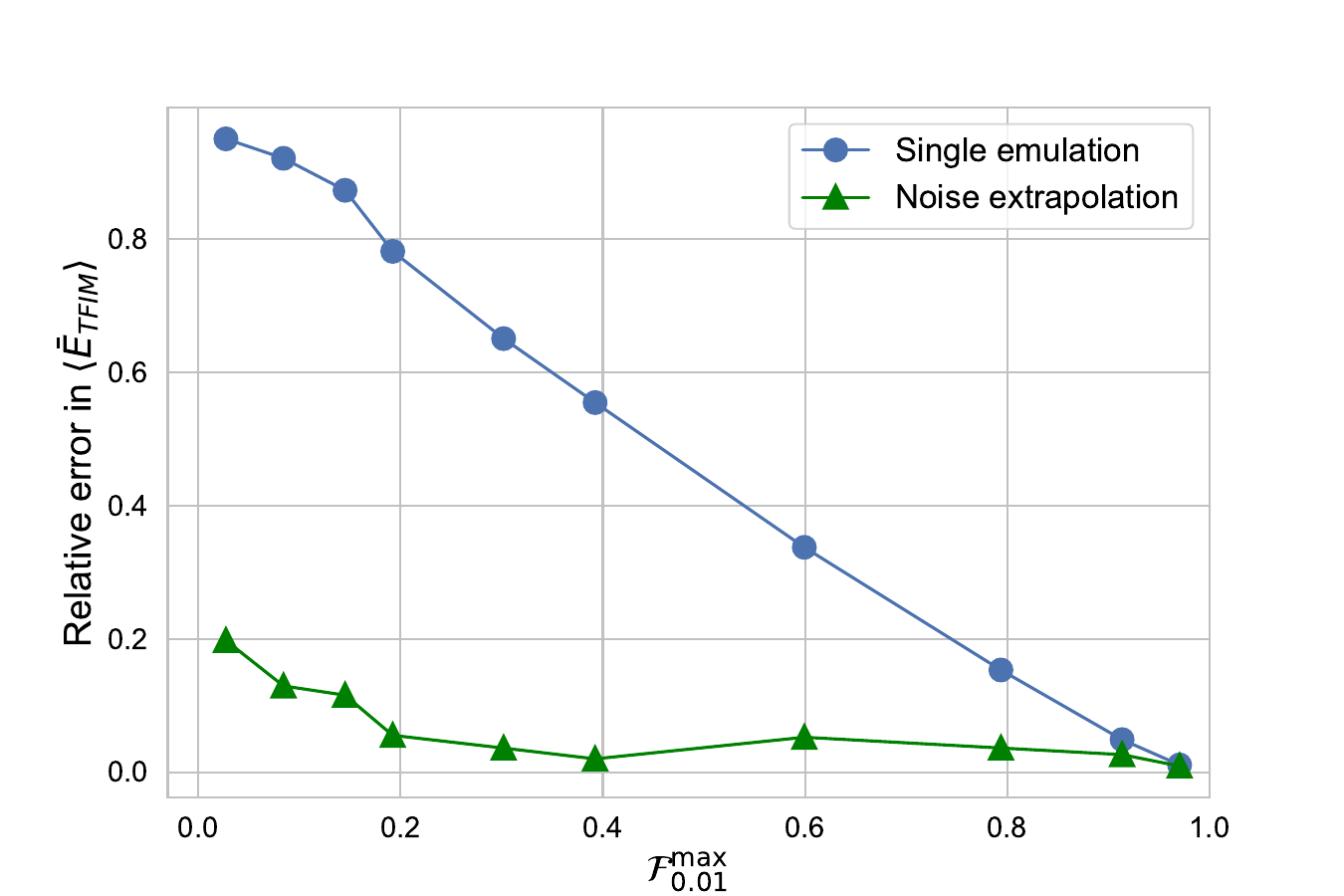}}%
    ~
    \subfigure{\label{fig:b}\includegraphics[width=0.5\textwidth]{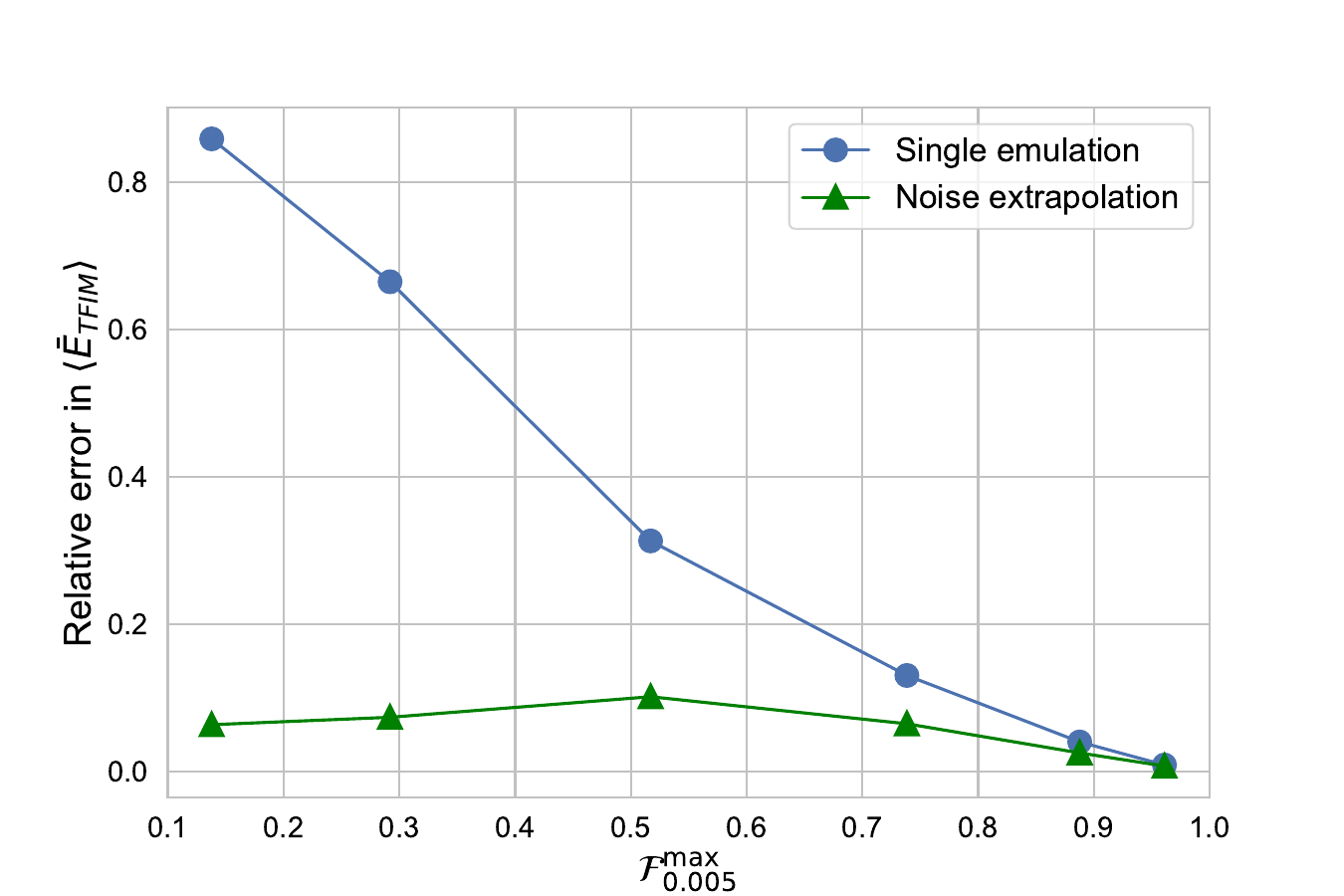}}
    \caption{Relative error in the energy per site $\ex{\bar{E}_{\text{TFIM}}}$ against emulation fidelity $\mathcal{F}_{\lambda^*}^{\max}$ at the target noise strength, for a circuit implementing 10 Trotter steps of time evolution of a $2 \times 7$ (14 qubit) instance of the TFIM under (a) depolarizing noise at $\lambda^*=0.01$ with exact value $\ex{\bar{E}_{\text{TFIM}}}_{\lambda^*} = -1.833$ and (b) cat noise at $\lambda^*=0.0005$ with exact value $\ex{\bar{E}_{\text{TFIM}}}_{\lambda^*} = -1.710$, for both a single $D_{\max}$ emulation at the target noise strength (blue line) and non-zero noise extrapolation using only $D \leq D_{\max}$ emulations (green line). Emulations were run with $D_{\max} \in \{8, \dots, 512\}$ in order to obtain different emulation fidelities between 0 and 1.}
    \label{fig:table_1}
\end{figure*}

We ran emulations for different values of $D_{\max}$ ($8, \dots, 512$). The results in Figures~\ref{fig:table_1}~(a) \& (b) show the relative errors in $\ex{\bar{E}_{\text{TFIM}}}$ for estimates obtained from single $D_{\max}$ emulations at $\lambda^*$ vs.~ones obtained through non-zero noise extrapolation (using only emulations with bond dimensions $D \leq D_{\max}$), against the emulation fidelity $\mathcal{F}_{\lambda^*}^{\max}$ for depolarizing and cat noise respectively. Moving to the right on the $x$-axes of Figures~\ref{fig:table_1}~(a) \& (b) corresponds to increasing $D_{\max}$, which increases $\mathcal{F}_{\lambda^*}^{\max}$ whilst shrinking the errors of both the single $D_{\max}$ emulation and the non-zero noise extrapolated expectation values. Here we observe (i) an advantage of non-zero noise extrapolation when only low-fidelity emulations are obtainable for the target noise strength, and (ii) that whilst this advantage shrinks as $\mathcal{F}_{\lambda^*}^{\max}$ increases,  non-zero noise extrapolation remains more accurate than the corresponding single $D_{\max}$ emulations. 

\subsubsection{Fermi-Hubbard model (FHM)}
\begin{figure*}[t!]
    \centering
    \subfigure{\includegraphics[width=0.8\textwidth]{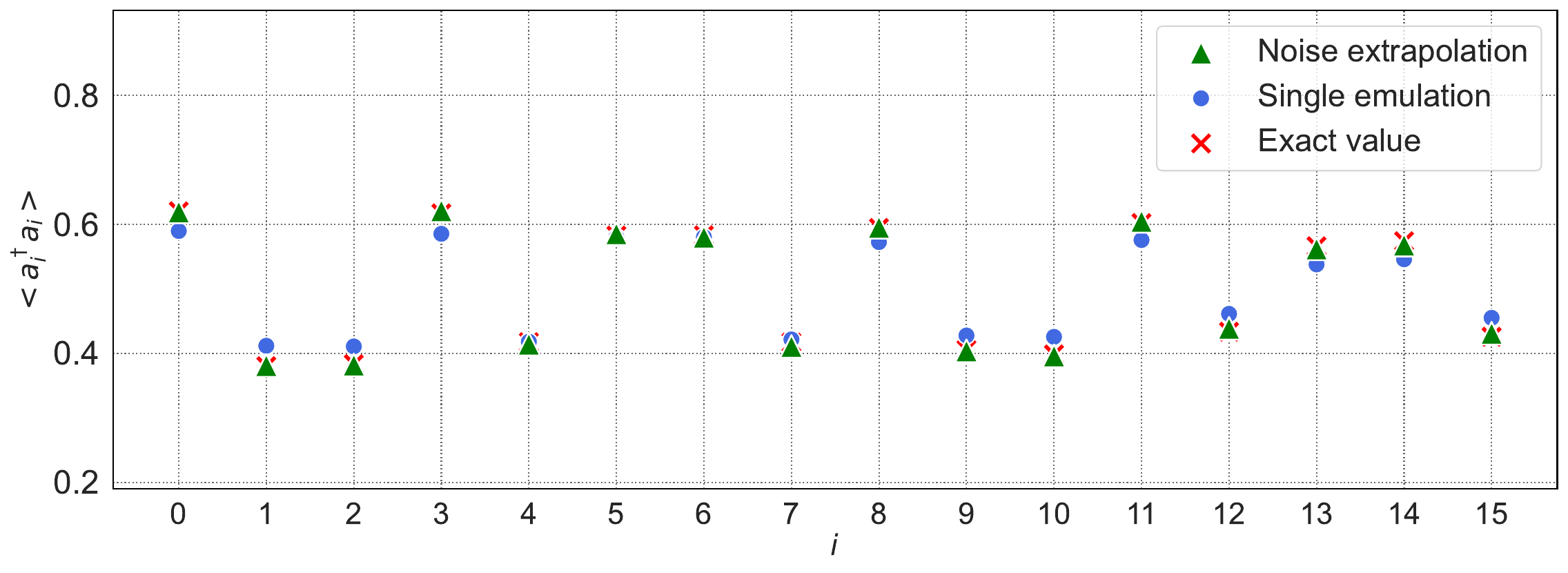}}
        
    \subfigure{\includegraphics[width=0.8\textwidth]{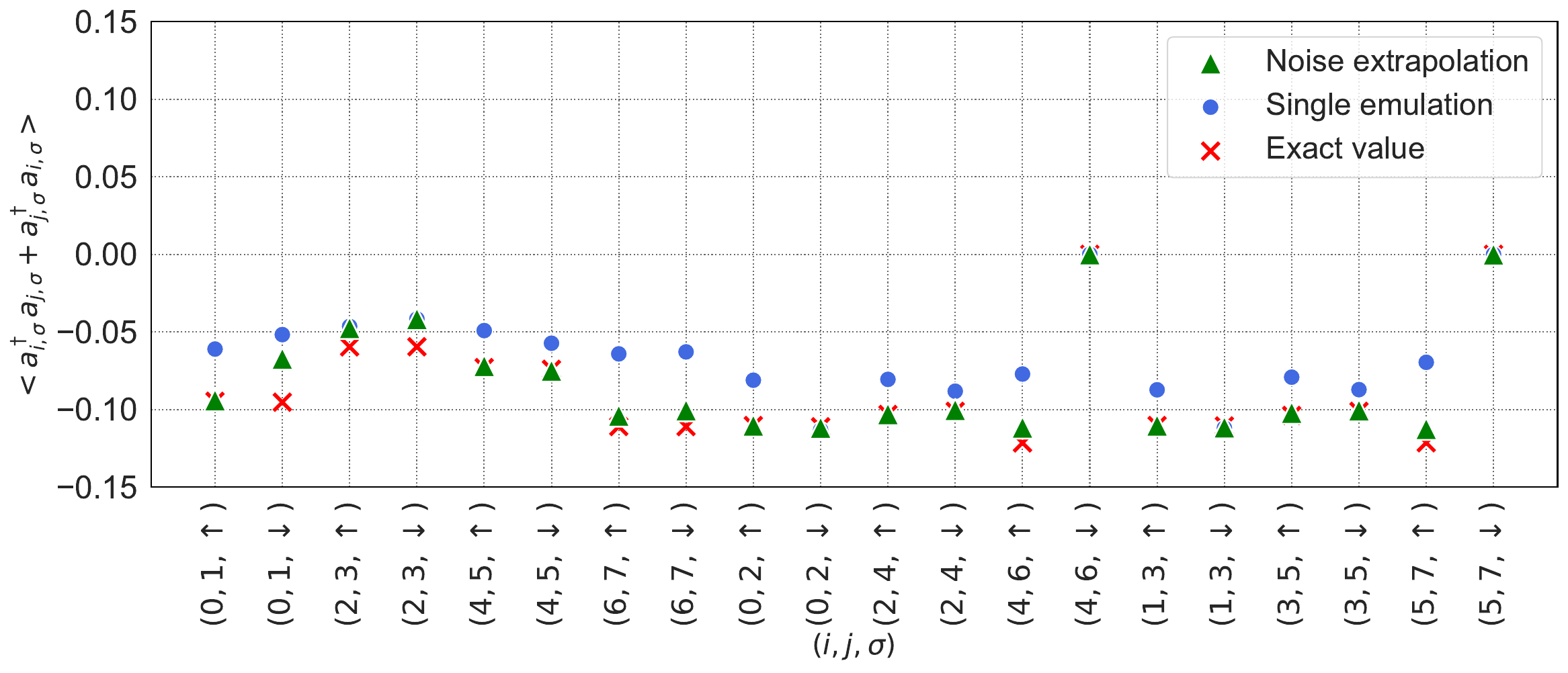}}
    
    \caption{Expectation values of (a) $a^\dagger_{i\, \sigma} a_{i, \sigma}$ and (b)  $(a^\dagger_{i,\sigma} a_{j,\sigma} + a^\dagger_{j,\sigma} a_{i,\sigma})$ after $10$ Trotter steps of a $2\times4$ FHM with $U/t=8$ and depolarising noise at $\lambda^*=0.001$. Blue circles: expectation values from a single emulation with $D_{\max}=64$ (and $\mathcal{F}_{\lambda^*}^{\max}=0.69$) yielding average absolute errors of $0.0214$ (a) and $0.0235$ (b) compared to the exact values (red crosses). Green triangles: expectation values computed via non-zero noise extrapolation with $D\leq64$, yielding average absolute errors of $0.0082$ (a) and $0.0072$ (b) compared to the exact values.}
    \label{fig:fhm_u/t=8_scatter}
\end{figure*}

We consider a 2x4 lattice for the Fermi-Hubbard model with open boundary conditions (which corresponds to 16 qubits because we have two qubits per site), and compute expectation values of all single-site number operators $\ad_{i,\sigma}a_{i,\sigma}$ and nearest-neighbour hopping terms $(\ad_{i, \sigma}a_{j,\sigma} + \ad_{j, \sigma}a_{i,\sigma})$ in their qubit (Jordan-Wigner) encodings (see Appendix~\ref{app:fhm}) after 10 Trotter steps (dt = 0.1) under depolarizing noise with target noise strength $\lambda^* = 0.001$, where the initial state is an anti-ferromagnetically aligned state on a 2D square lattice (see Appendix~\ref{app:fhm} for further details on this circuit and choice of initial state). The sites in the 2x4 lattice are row-major ordered (i.e.~the first row contains sites 0 and 1, the second 2 and 3, the third 4 and 5, and the last 6 and 7).
We use a maximum bond dimension of $D_{\max} = 64$, which yields $\mF_{\lambda^*}^{\max} = 0.69$ while $\mF_{\lambda}^{\max} \geq 0.99$ for $\lambda \geq 0.03$. Figures~\ref{fig:fhm_u/t=8_scatter} (a) \& (b) show the expectation values for the number operators and hopping terms, respectively. Compared to the corresponding Figure~\ref{fig:dep_ising_14_scatter} for the TFIM, the accuracies of expectation values coming from single $D_{\max}$ emulations are higher, but we still see an appreciable improvement in accuracy from non-zero noise extrapolation.

\begin{figure}
    \centering
    \includegraphics[width=0.6\linewidth]{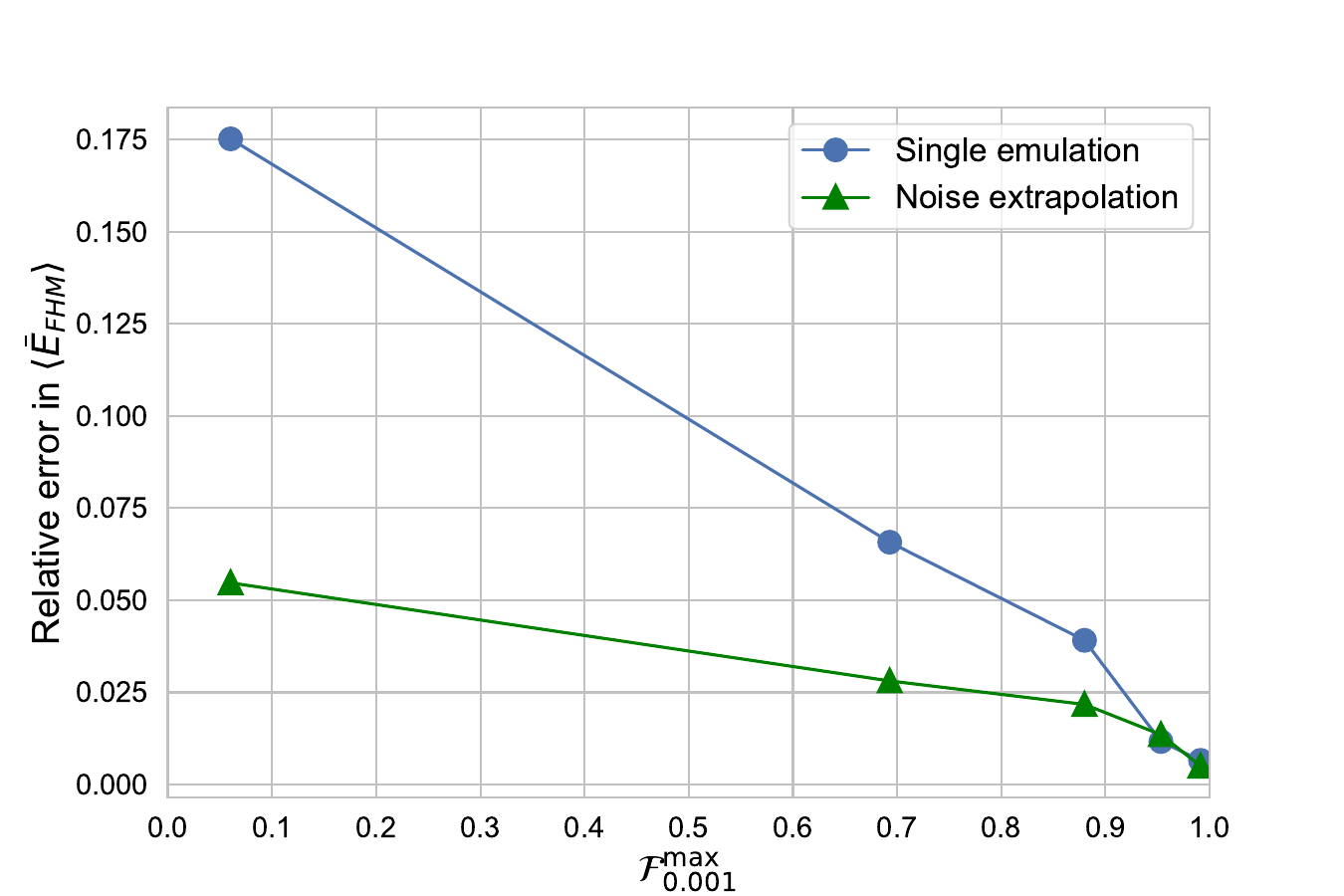}
    \caption{Relative error in the energy per site $\ex{\bar{E}_{\text{FHM}}}$ (with exact expectation value $\ex{\bar{E}_{\text{FHM}}}_{\lambda^*} = 1.6789$) against emulation fidelity $\mathcal{F}_{\lambda^*}^{\max}$, for a circuit implementing 10 Trotter steps ($dt=0.1$) of time evolution of a  $2 \times 4$ (16 qubit) instance of the FHM under depolarizing noise at $\lambda^*=0.001$, for both a single $D_{\max}$ emulation at the target noise strength (blue line) and non-zero noise extrapolation using only $D\leq D_{\max}$ emulations (green line). Emulations were run with $D_{\max}\leq 512$  to obtain different emulation fidelities between 0 and 1.}
    \label{fig:table_2}
\end{figure}

To quantify the improvement in accuracy provided by non-zero noise extrapolation, we again consider the observable $\bar{E}_{\text{FHM}}$ corresponding to the energy per site of the Fermi-Hubbard model,
\[
    \bar{E}_{\text{FHM}} := \frac{1}{n} H_{\text{FHM}}\,,
\]
where $n$ is the number of sites and $H_{\text{FHM}}$ is the Hamiltonian defined in Appendix~\ref{app:fhm}. Once again, to compute $\ex{\bar{E}_{\text{FHM}}}$ we use the non-zero noise extrapolated values for the hopping and on-site terms of $H_{\text{FHM}}$. In Figure~\ref{fig:table_2} we show the relative error in $\ex{\bar{E}_{\text{FHM}}}$ for an estimate obtained from a single $D_{\max}$ emulation at $\lambda^*=0.001$ vs.~one obtained through non-zero noise extrapolation (using bond dimensions $D_{\max} \in \{8, \dots, 512\}$), against the emulation fidelity $\mathcal{F}_{\lambda^*}^{\max}$. As with the TFIM, we observe (i) a clear advantage to using non-zero noise extrapolation when $\mathcal{F}_{\lambda^*}$ is low, and (ii) that this advantage shrinks with increasing emulation fidelity $\mathcal{F}_{\lambda^*}$, but that the non-zero noise extrapolated values remain more accurate than the corresponding single $D_{\max}$ emulation expectation values.

This is a good moment to discuss the added runtime due to the extra simulations required for non-zero noise extrapolation compared to simply increasing the bond dimension. Given that entanglement entropy scales as the logarithm of the bond dimension $D$, one typically needs to double the bond dimension to significantly improve the accuracy of a single emulation. The runtime of MPS time-evolution scales as $D^3$, and therefore a doubling of bond dimension corresponds to a factor 8 increase in runtime. Non-zero noise extrapolation also requires of the order of 8 separate simulations to be accurate. Therefore, non-zero noise extrapolation using simulations of bond dimension $D$ takes roughly as much time as a single emulation at bond dimension $2D$, making non-zero noise extrapolation faster whenever more than a doubling of bond dimension is needed for a single emulation to achieve the same accuracy. From Figures~\ref{fig:table_1} and~\ref{fig:table_2}, where the maximum bond dimension doubles from one data point to the next, we see that in both cases a low bond dimension non-zero noise extrapolated result is more accurate than a single emulation of twice that bond dimension (until we reach the regime where single emulations are almost exact).\footnote{There is also some overhead from the lower $D$ simulations required to do bond dimension extrapolation at fixed noise strengths, but due to the $D^3$ scaling, these simulations are sub-leading in terms of runtime.} In addition to the above, it should be noted that in practice compute is typically memory restricted (e.g. fixed-memory GPUs), and therefore allows for emulations up to a fixed bond dimension. That is, it is typically the memory, not the runtime, that forms the bottleneck in (GPU-accelerated) MPS time-evolution emulations. The main practical advantage of non-zero noise extrapolation is therefore in its ability to obtain more accurate simulations from a fixed amount of computational memory.

\subsection{Large systems}\label{sec:large-systems}
We now turn our attention to noisy circuits on large numbers of qubits, in order to analyse the behaviour of non-zero noise extrapolation for systems far beyond the full-state regime. We begin by assessing the accuracy for a 60-qubit circuit implementing time evolution of the XY model (XYM) in one dimension, where we can obtain approximations of the true expectation values via exact trajectory simulations of matchgate circuits -- see Appendix~\ref{app:matchgate}. 

After having benchmarked non-zero noise extrapolation on the 60-qubit XYM, we move into the territory where exact simulations are not possible, and investigate the performance of our method there. This involves the estimation of expectation values of observables for noisy emulations of a 60-qubit FHM instance. For these emulations we used Fermioniq's tensor network emulator, Ava, at close to maximum bond dimension given the RAM available: up to $5000$ for the pure-state noiseless emulation, and up to $1500$ for the mixed-state noisy emulations. We present these results as an example of how the non-zero noise extrapolation method combined with tensor network emulation can be utilised to study the effects of noise on quantum circuits out of reach of existing methods. For this system we achieve emulation fidelities comparable to those obtained in Section~\ref{sec:results_exact}, which gives confidence in the reliability of the method in this regime. This confidence is further backed up by a qualitative assessment of our results.

\subsubsection{$XY$ model on 60 qubits}\label{sec:results_large_xy}
Here we consider a 1-dimensional $XY$ model on 60 qubits, which allows us to compare the results of non-zero noise extrapolation with an approximation of the true expectation values obtained via (exact) trajectory simulations over matchgate circuits (see Appendix~\ref{app:matchgate}). Since 1-dimensional circuits are more amenable to emulation via tensor networks, the bond dimension was limited to 300 for the mixed-state runs and 100 for the pure-state runs in order to keep the fidelities down and give the non-zero noise extrapolation more work to do. Figure~\ref{fig:fidelities_YX_30,31} shows the emulation fidelities obtained for these bond dimensions for different strengths of the noise.

\begin{figure*}[htb!]
    \centering
    \includegraphics[width=0.75\linewidth]{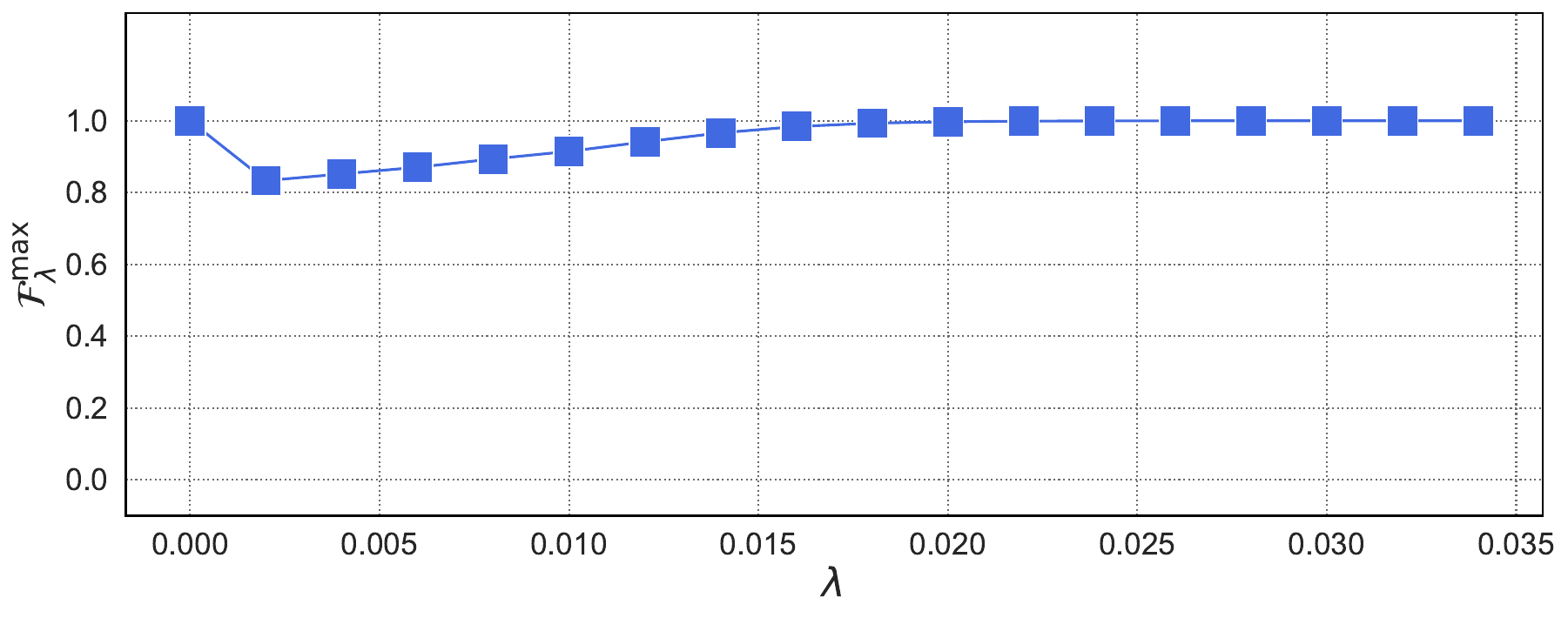}
    \caption{Emulation fidelity $\mathcal{F}_\lambda^{\max}$ against noise strength $\lambda$, for a circuit implementing 30 Trotter steps of time evolution of the 1-dimensional $XY$ model on 60 qubits (see Appendix~\ref{app:xym}). For the pure-state emulation at $\lambda=0$ we use $D_{\max} = 100$ and for the mixed-state emulations at $\lambda>0$ we use $D_{\max} = 300$.}
    \label{fig:fidelities_YX_30,31}
\end{figure*}

Note that, in addition to the trajectory simulations, we also obtained high fidelity tensor network emulations at the target noise strength ($\mathcal{F}_{0.002} = 0.9891$) by using larger bond dimension ($D=800$), which we could also have used as a point of comparison. However, trajectory sampling comes with a different and contrasting source of error to tensor network emulations, and from it  we can also obtain a quantitative estimate for the standard deviation and thus a reliable estimate for the error without having to rely on the heuristic fidelity of a tensor network emulation. Note that, although we (and others in the literature, e.g. \cite{Clark_dm}) have found the heuristic fidelity of a tensor network emulation to be a good guide to the accuracy of the emulation, there is no mathematically rigorous bound on the size of the errors of expectation values for all observables. We have, therefore, chosen to compare the non-zero noise extrapolation method to trajectory sampling over matchgate circuits for the XY model as a sanity check.
 
\begin{figure*}[htb!]
    \centering
    \subfigure{\label{fig:a}\includegraphics[width=0.5\textwidth]{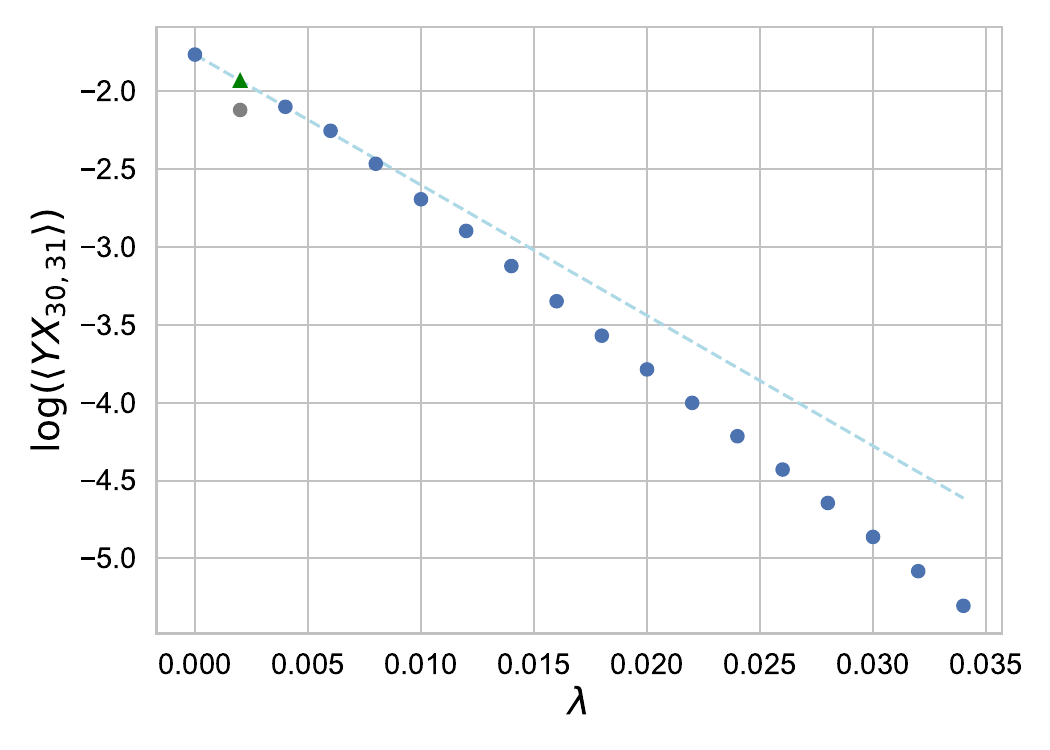}}%
    ~
    \subfigure{\label{fig:b}\includegraphics[width=0.5\textwidth]{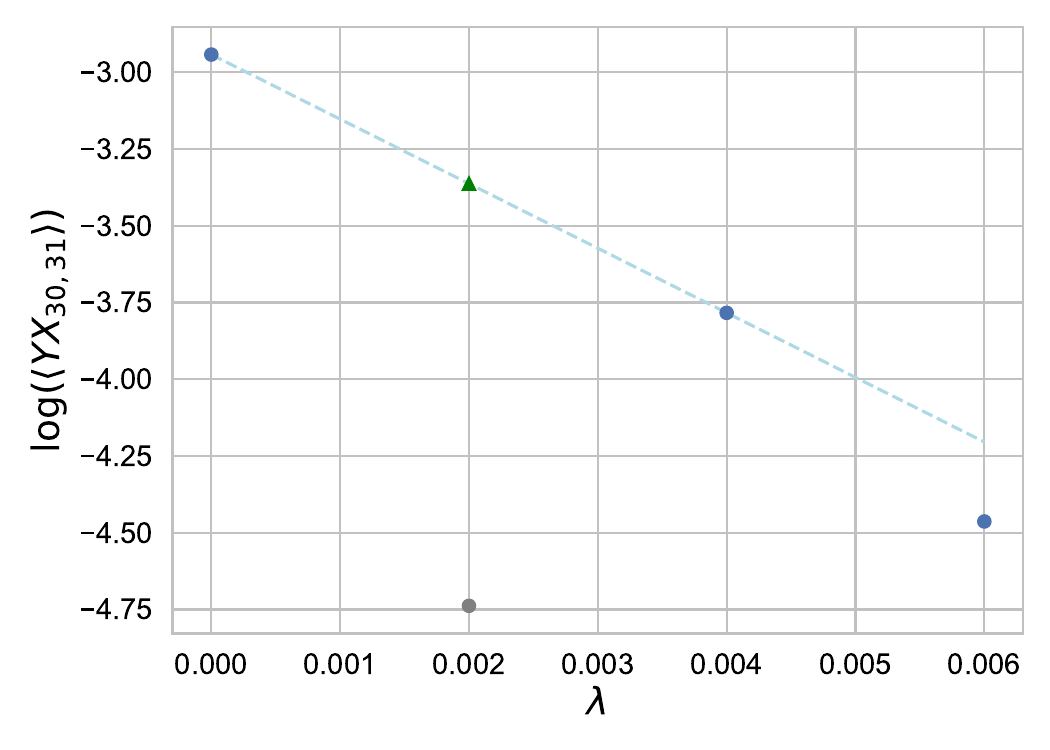}}
    \caption{Log of $\left|\ex{YX_{30,31}}\right|$ (left) and $\left|\ex{Z_{30}}\right|$ (right) obtained via fidelity extrapolation for simulations at different noise strengths $\lambda$, with a weighted straight line fit (dashed blue line) ignoring the anomalous $\lambda=0.002$ point (gray circle) and corresponding extrapolated (log) expectation value at the target noise strength $\lambda^* = 0.002$ (green triangle). Same circuit and emulation settings as for Figure~\ref{fig:fidelities_YX_30,31}.}
    \label{fig:XYM_expvals}
\end{figure*}

In Figure~\ref{fig:XYM_expvals}(a) we show the log of the absolute values~\footnote{In this case all values were negative, hence the need to take the absolute value before computing the log. In general, criteria~\ref{crit:sign} from Section~\ref{ssec:implementation} ensures that the observables $\langle O\rangle_\lambda$ will all have the same sign.} of a two-site observable from the middle of the chain $\ex{YX}_{30,31}$ obtained after fidelity extrapolation for each noise strength $\lambda$. It is clear that the data point for the target nose strength, which we choose to be $\lambda^*=0.002$ (shown as a gray circle in the plot) is an outlier, so we ignore this point when performing the non-zero noise extrapolation~\footnote{Upon further inspection, we could see that the extrapolation for this data point was quite poor. This suggests that a tightening of our convergence criterion~\ref{ssec:implementation} \ref{crit:conv} would take care of automatically removing this point.}. The extrapolation was performed via a weighted fit as described in Section~\ref{ssec:implementation}, using the same settings as for all other simulations, and yielded an estimate of $\overline{\ex{YX}}_{30,31} = -0.1450$.

In Figure~\ref{fig:XYM_expvals}(b) we show the log values of a single-site observable from the middle of the chain $\ex{Z}_{30}$, obtained after fidelity extrapolation for each noise strength. Here, fewer data points were retained by the criteria in Section~\ref{ssec:implementation} (\ref{crit:conv}-\ref{crit:sign}), which was due to the fact that the value of $\ex{Z}_{30}$ changed signs with increasing noise strength. Once again, we see that the data point for $\lambda^* = 0.002$ is a clear outlier, and we ignored this data point when performing the extrapolation. The results of the extrapolation yielded an estimate of $\overline{\ex{Z}}_{30} = -0.0346$. Since so few data points pass the criteria in Section~\ref{ssec:implementation} (\ref{crit:conv}-\ref{crit:sign}), it is challenging to perform a meaningful extrapolation for the case of $\langle Z_{30}\rangle$. Therefore, in Appendix~\ref{ssec:xym_results} we also perform a fit to the exponential Ansatz~\eqref{eqn:exp_Ansatz} directly, and find extrapolated values which are close to those that we have presented here in the main text.

\begin{figure*}[t!]
    \centering
    \subfigure{\label{fig:a}\includegraphics[width=0.5\textwidth]{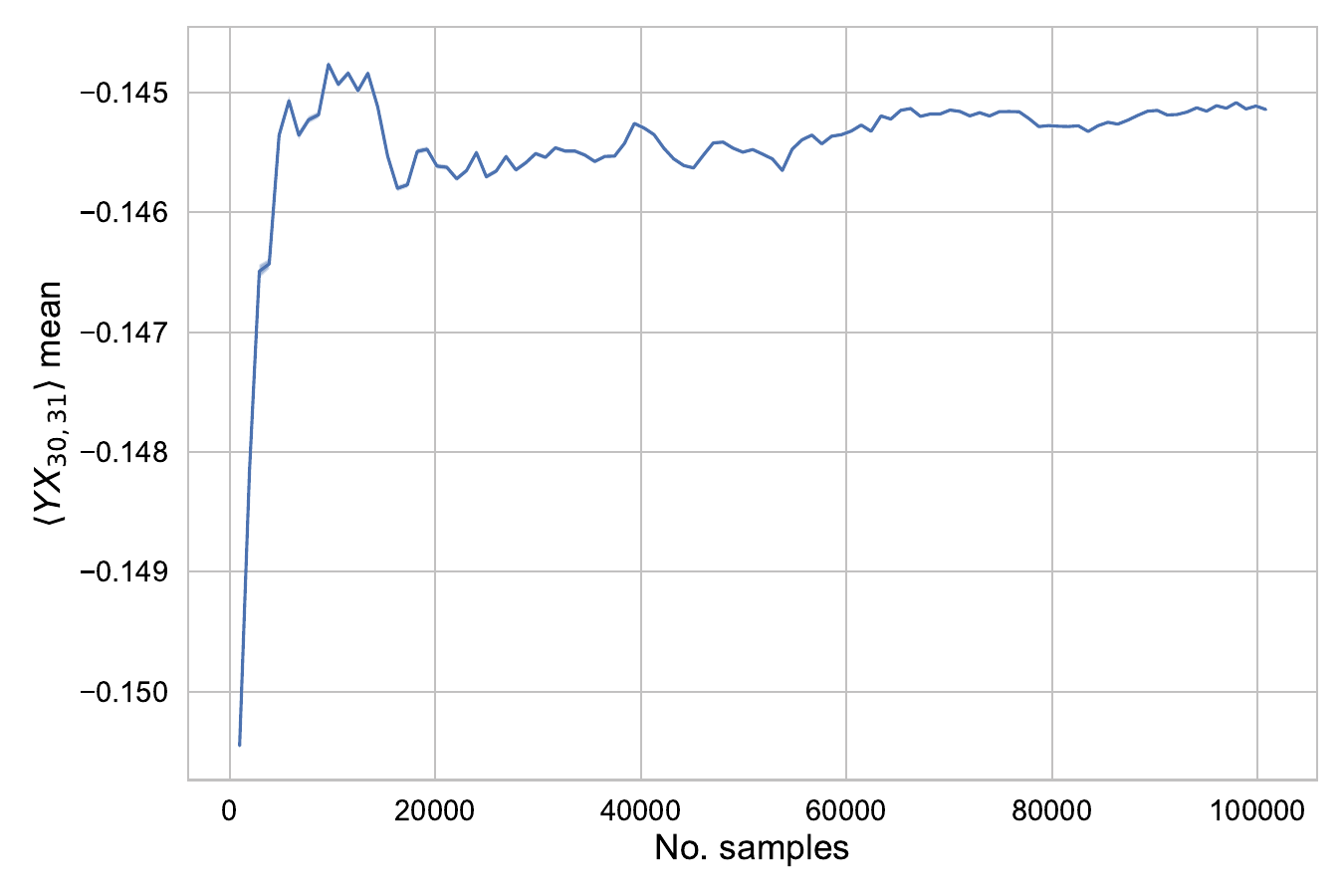}}%
    ~
    \subfigure{\label{fig:b}\includegraphics[width=0.5\textwidth]{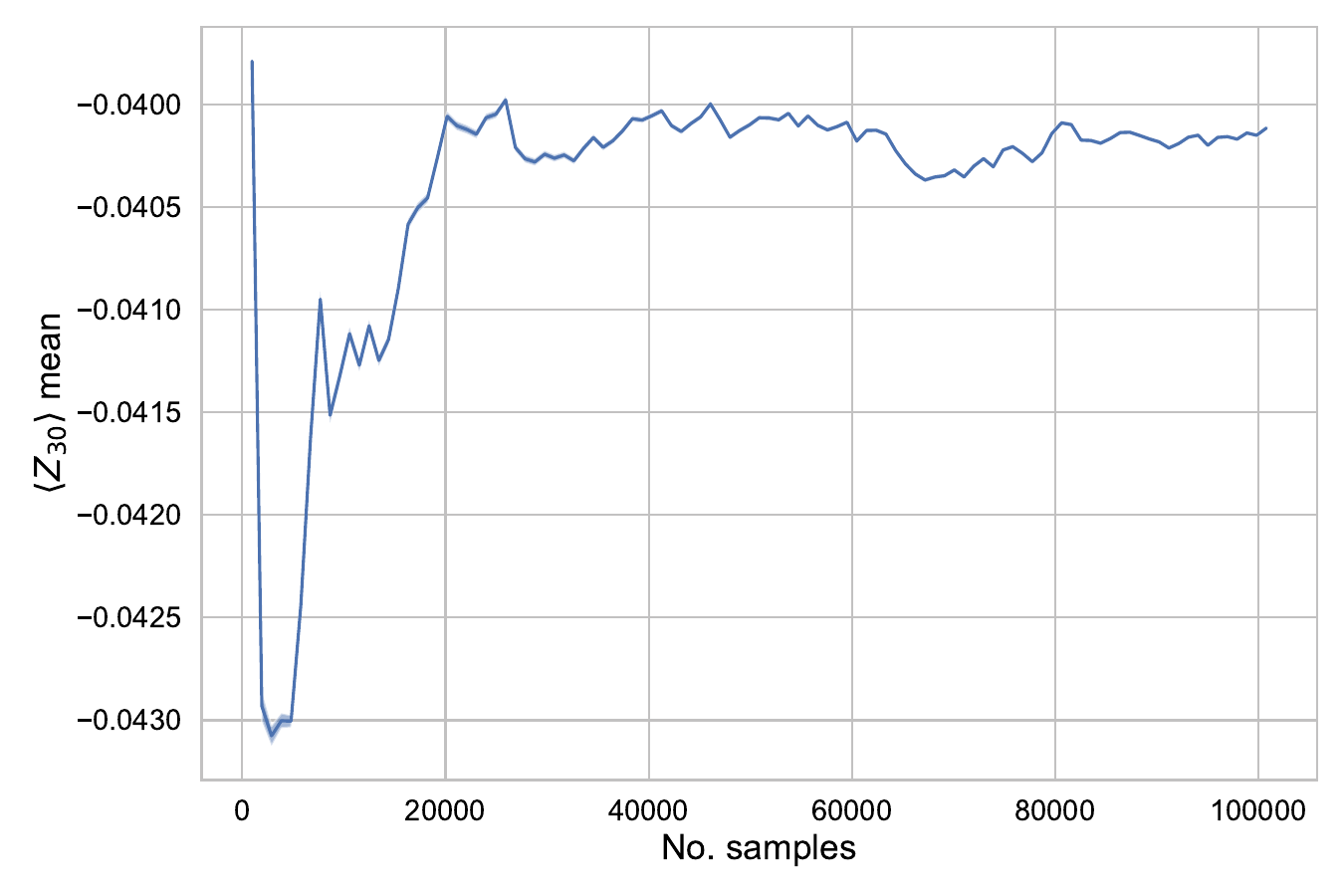}}
\caption{Further details on matchgate simulation by trajectory sampling. Mean value of (a) $\ex{YX_{30,31}}$ and (b) $\ex{Z_{30}}$  as a function of the number of sampled trajectories of the noisy circuit (see Appendix~\ref{app:matchgate}). After $100,800$ samples, the sample mean of $\ex{YX_{30,31}}$ was $-0.1451$ (with a standard deviation of 0.0032) and the sample mean of $\ex{Z_{30}}$ was $-0.04011$ (with a standard deviation of 0.0028).}
\label{fig:matchgate_trajectory_expvals}
\end{figure*}

To evaluate the accuracy of the noise extrapolation, we estimated the values of $\ex{YX}_{30,31}$ and $\ex{Z}_{30}$ at noise strength $\lambda^* = 0.002$ via trajectory simulations of the corresponding matchgate circuit. Sampling 100,800 trajectories, each of them an exact pure-state simulation, resulted in mean expectation values of $-0.1451$ (standard deviation $0.0032$) and $-0.0401$ (standard deviation $0.0028$), respectively.
Figure~\ref{fig:matchgate_trajectory_expvals} shows the convergence behaviour of the means for both observables. In comparison with the noise-extrapolated values, we can see close agreement for $\ex{YX}_{30,31}$ and slightly poorer agreement for $\ex{Z}_{30}$. This is likely due to the more difficult non-zero noise extrapolation for $\ex{Z}_{30}$ that we saw above due to fluctuating sign of the high-noise data points. How to handle such cases could be an interesting direction for future research. Table~\ref{tab:xy_summary} summarises the results arising from single emulations, non-zero noise extrapolation, and trajectory simulations for this circuit. 

\begin{table*}
\centering
\begin{tabular}{l|l|l|l}
\multicolumn{1}{c|}{Observable} &
  \multicolumn{1}{c|}{Trajectory simulation} &
  \multicolumn{1}{c|}{\begin{tabular}[c]{@{}c@{}}Single emulation \\ ($D=300$)\end{tabular}} &
  \multicolumn{1}{c}{\begin{tabular}[c]{@{}c@{}}Non-zero noise extrapolation \\ ($D=300$)\end{tabular}} \\ \hline
$\ex{YX}_{30,31}$ &
  -0.1451 &
  -0.1201 (0.1723) &
  -0.1450 (0.0007) \\ \hline
$\ex{Z}_{30}$ &
  -0.0401 &
  -0.0088 (0.7805) &
  -0.0346 (0.1372) \\ \hline
\end{tabular}
\caption{Expectation values approximated by the trajectory method, by fidelity extrapolation of single emulations at $D_{\max}=300$, and by non-zero noise extrapolation at $D_{\max}=300$. The figures in brackets are the relative errors compared to the trajectory estimate.}
\label{tab:xy_summary}
\end{table*}

\subsubsection{Fermi-Hubbard model on 60 qubits}
\label{ssec:fhm_large_results}
\begin{figure*}[htp]
\centering
\includegraphics[width=0.65\textwidth]{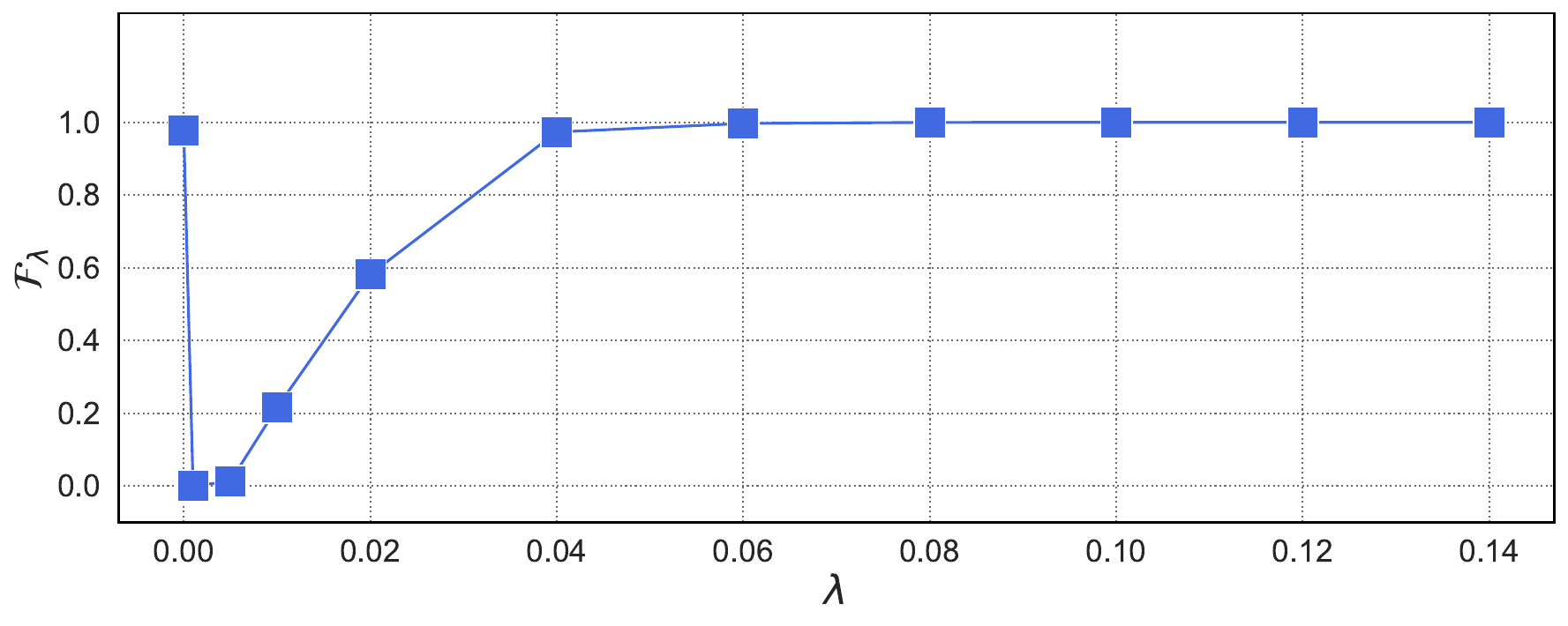}
\caption{Emulation fidelity $\mathcal{F}_\lambda^{\max}$ obtained at $D_{\max} = 5000$ for $\lambda = 0$ and $D_{\max} = 1500$ for $\lambda > 0$ vs. noise strength $\lambda$ for emulations of circuits implementing 10 Trotter steps of time evolution of a $5 \times 6$ (60 qubits) instance of the FHM with $U/t = 8$ (see Appendix~\ref{app:fhm}).}
\label{fig:fidelities_large_systems_fhm}
\end{figure*}

We emulated noisy circuits implementing time evolution of a $6 \times 5$ ($60$ qubit) FHM instance for 10 Trotter steps ($dt=0.1)$, with $U/t = 8$. The initial state is an anti-ferromagnetically aligned product state on a $2D$ square lattice (see Appendix~\ref{app:fhm}). The emulation fidelities for each noise strength are shown in Figure~\ref{fig:fidelities_large_systems_fhm}.

\begin{figure*}[htb!]
    \centering
    \includegraphics[width=0.65\linewidth]{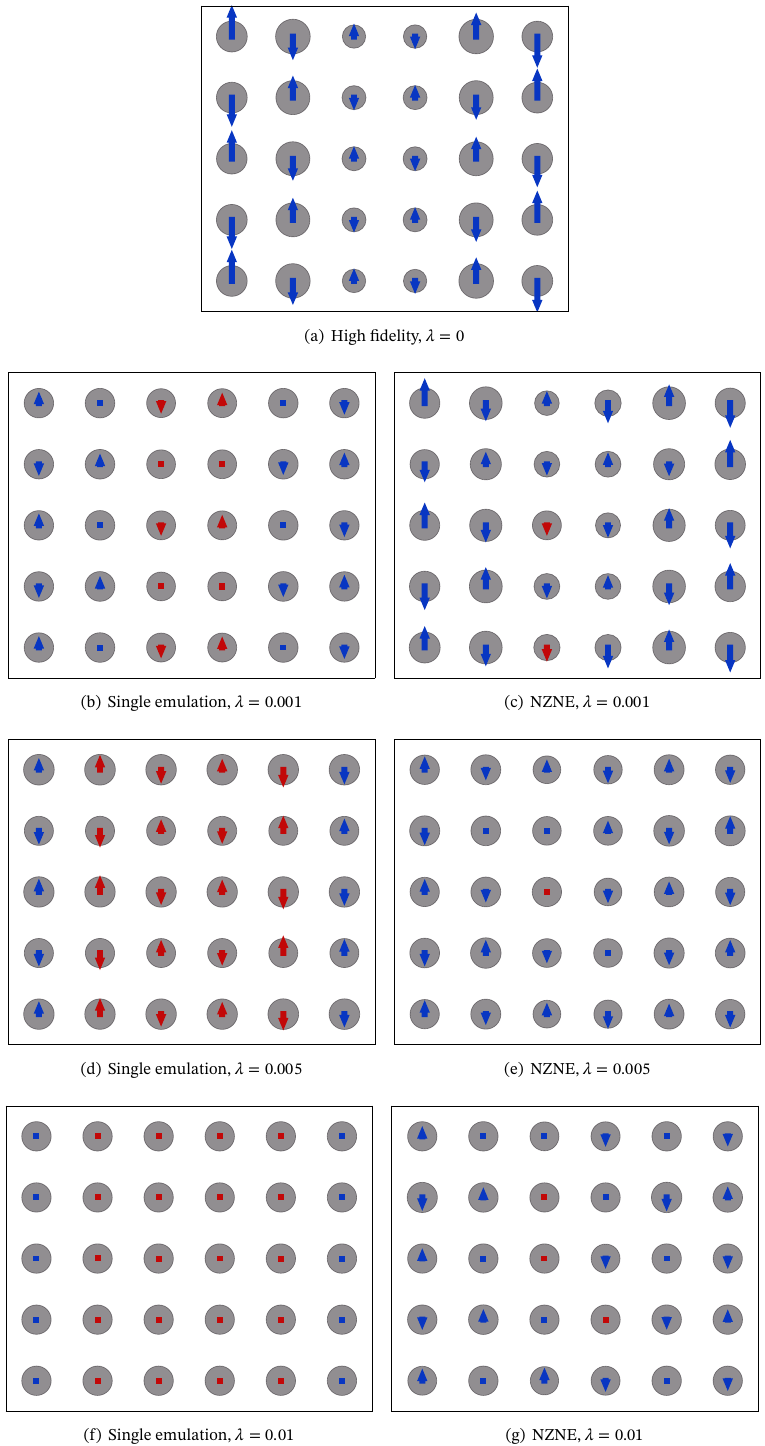}
\caption{Filling (gray circles) and magnetization (red/blue arrows) per site of a $5 \times 6$ (60 qubit) instance of FHM after 10 Trotter steps  ($dt=0.1$), at $U/t=8$. (a) Noiseless ($\lambda=0$) case. (b) - (g) Filling and magnetization per site for increasing noise strengths (top to bottom), when using results from a single emulation plus fidelity extrapolation (left) vs.~using non-zero noise extrapolation (right). Blue arrows show agreement with the sign of the noiseless magnetization, red arrows show disagreement.} 
    \label{fig:large_quiver}
\end{figure*}

\begin{figure*}[htb!]
    \begin{minipage}[t!]{\linewidth}
    \centering
    \includegraphics[width=0.9\linewidth]{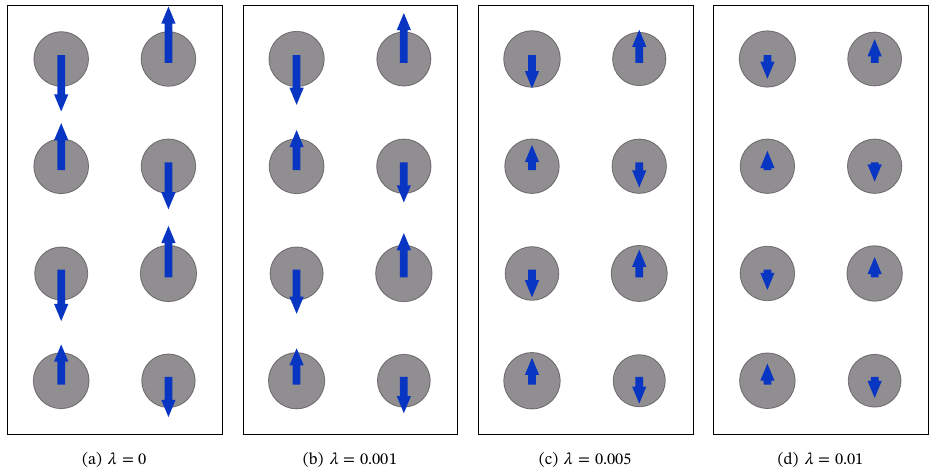}
    \caption{Filling (gray circles) and magnetization (red/blue arrows) per site of a $2 \times 4$ (16 qubit) instance of FHM after 10 Trotter steps, at $U/t=8$. (a) Noiseless ($\lambda=0$) case. (b) - (d) Filling and magnetization per site for increasing noise strengths. All results come from a very high fidelity ($>0.999$ emulation). Blue arrows show agreement with the sign of the noiseless magnetization, red arrows show disagreement (of which there is none).} 
    \label{fig:small_quiver}
    \end{minipage}
\end{figure*}

To demonstrate the utility of our method for this system, we consider two commonly measured observables: the filling and magnetization per site, given by the observables $n_i = \left(\hat{n}_{i\uparrow} + \hat{n}_{i\downarrow}\right) $ and $m_i = \left(\hat{n}_{i\uparrow} - \hat{n}_{i\downarrow}\right)$, respectively. Under the JW encoding, these are computed from Pauli $Z$ observables on the qubits (see Appendix~\ref{app:fhm}), which are measured at the end of the circuit. We now verify that using non-zero noise extrapolation to estimate the expectation values of these observables yields qualitatively sensible results, in the absence of exact data to compare against. We consider circuits subject to depolarizing noise of strengths $0.001$, $0.005$, and $0.01$ (corresponding to two-qubit gate fidelities of $0.9995$, $0.9975$, and $0.995$ respectively, assuming the one-qubit gates are noiseless). At these noise strengths, accurate emulations are out of reach of the tensor network emulator (as can be seen in Figure~\ref{fig:fidelities_large_systems_fhm}), and the noise strength of $0.001$ in particular is a very challenging regime. Nevertheless, non-zero noise extrapolation can be used to provide (what appear to be) reliable improvements in accuracy for these emulations.

We visualise the data as `quiver' plots in Figure~\ref{fig:large_quiver}, in which we show the zero-noise data (obtained via a high-fidelity ($\mathcal{F}_0 = 0.9775$) pure-state emulation at $D=5000$) as a point of comparison to the data obtained for circuits subject to depolarizing noise. We can see that without noise, the spins of neighbouring sites tend to anti-align, with filling of sites on the left and right edges higher than those in the centre. Below this we show the data obtained from the noisy circuits: from a single emulation at the target noise strengths (left plots), compared to the non-zero noise extrapolated data (right plots). The colours in the plots indicate qualitative agreement with the pure-state data: blue arrows show magnetizations with agreeing signs, and red disagreeing signs. We see that the non-zero noise extrapolated data tends to better match the patterns of the noiseless data, whereas the data coming from single emulations quickly loses the anti-alignment properties. In both cases, we find that increasing noise strengths lead to more uniform fillings and dampened magnetization magnitudes. 

To verify that this is `expected' behaviour, we can compare to high-fidelity data for smaller systems. In Figure~\ref{fig:small_quiver}, we show the same observables for a smaller $2 \times 4$ (16 qubit) FHM instance, obtained via very high-fidelity ($\mathcal{F} > 0.999$) tensor network emulations. We again display the noiseless data alongside the noisy results, with the same strengths of depolarizing noise as above ($0.001$, $0.005$, $0.01$). Here we see the same trend exhibited by the non-zero noise extrapolated data for the larger system: dampened magnetization magnitudes of the same sign, with dampening becoming stronger with increased noise strength, suggesting that non-zero noise extrapolation is improving the quality of results for the 60-qubit system discussed above.

Of course these are not one-to-one comparisons: noise appears to have a smaller effect on the small system than the large one, which is expected since the effects of noise scales both with circuit depth \textit{and} number of qubits. For completeness, we include all the data used to generate these plots and additional details on how it was obtained in Appendix~\ref{app:further_hubbard_results}.

\section{Discussion and further directions}\label{sec:discussion}

In this paper we introduced the technique of non-zero noise extrapolation: a method for accurately estimating expectation values of observables from emulations of noisy quantum circuits when accurate emulations are not possible at the desired noise strength, but can be obtained at stronger ones. We benchmarked this method using Fermioniq's Ava for the tensor network emulations, and found that the method significantly improved the accuracy of expectation values within the regime that we could compare to exact results. We then demonstrated the application of non-zero noise extrapolation beyond this regime, applying it in particular to the computation of physical observables for the Fermi-Hubbard model on 60 qubits, subject to two-qubit depolarizing noise at target noise strengths that corresponds to two-qubit gate fidelities at the upper end of abilities of current quantum hardware.

\begin{figure*}[htb!]
    \centering
    \includegraphics[width=0.9\linewidth]{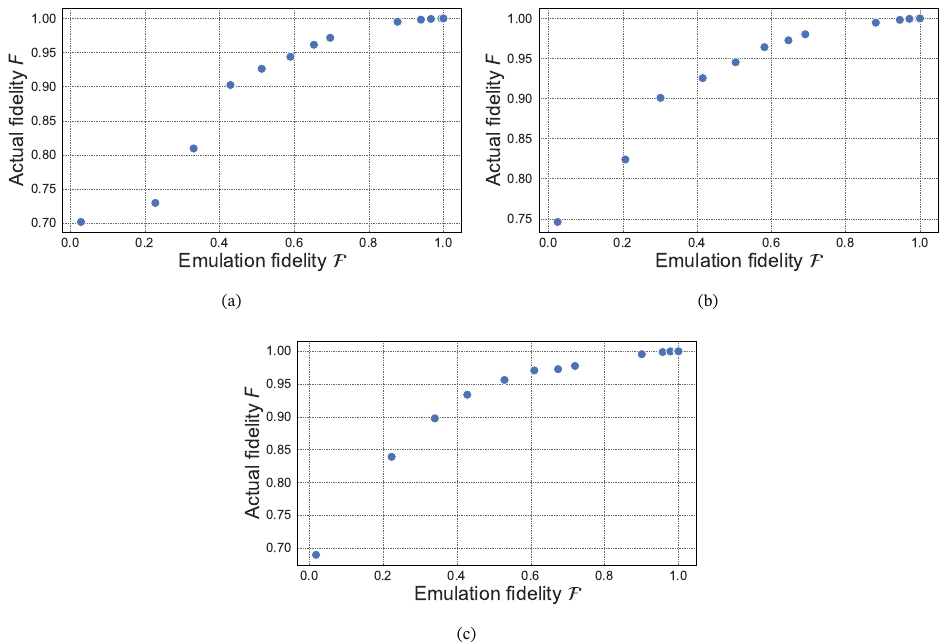}
    \caption{Emulation fidelity $\mF$ vs actual fidelity of the emulation computed via exact simulation. Each plot corresponds to emulations of the same circuit with different depolarizing noise strengths ((a): $\lambda=0.01$, (b): $\lambda=0.001$, (c): $\lambda=0$). The circuit used was time evolution of a 10-qubit TFIM instance for 10 Trotter steps with the settings given in Appendix~\ref{app:tfim}.}
    \label{fig:emulation-fid-vs-actual-fid}
\end{figure*}

\paragraph{Simulation of open quantum systems} 
A natural application of the technique we have proposed here would be to study open quantum systems, beyond quantum circuits. This could be achieved in a system with tuneable Lindbladian evolution, where a parameter quantitatively controls the decoherence rate of the evolving system.

\paragraph{Extension to 2d Tensor Networks} We have, so far, only applied this method to MPS based tensor network representations of the density matrix. However, it is natural to expect that the method proposed here could also be extended to representations of a density matrix using a 2d tensor network Ansatz (e.g. PEPO)~\cite{verstraete2004, Tindall2023, Zaletel2020}.

\paragraph{Extension to `non-parametrisable' noise models} In this work we considered noise models parametrised by a noise strength $\lambda$, in which the non-zero noise extrapolation was performed. This restriction is not necessary, and the extrapolation could be performed in an alternative way via techniques such as unitary folding~\cite{giurgica2020digital}. The algorithm could straightforwardly be adapted to such a setting, and would presumably work well so long as the noise decreases the entanglement built up by the circuit, and affects the expectation values of local observables in a predictable way. 

\paragraph{Optimisation of fitting procedure} Our approach to performing the extrapolation, including the criteria and loss function explained in Section~\ref{ssec:implementation}, were primarily developed by (limited) experimentation on specific examples. It would be valuable to optimise this approach further and to understand how generally it can be applied.

\noindent \textbf{Acknowledgments} -
We would like to thank Norbert Schuch for suggesting to use Matchgate circuits in order to benchmark the method for larger systems beyond the statevector regime. AT would like to thank Profesor Stephen Clark for a number of helpful conversations on sources of error from tensor network approximations of vectorised density matrices. AT acknowledges support from UK Engineering and Physical Sciences Research Council
(EP/SO23607/1). Finally, we would like to thank the entire team at Fermioniq, and especially Boris Ponsioen, Hjalmar Lindstedt, and Eline Welling for their input and co-development of Ava. This work was supported by the Dutch National Growth Fund (NGF), as part of the Quantum Delta NL Programme.

\section*{Data Availability}
Supporting data for this publication is openly available at~\cite{github2025a}, open source code developed for this project for the simulation of matchgate circuits is available at~\cite{github2025b}.

\appendix

\section{Further details on tensor network simulation of density matrices}

\subsection{MPO entanglement entropy}
\label{sec:mpo_entanglement_entropy}
As described in Section~\ref{sec:nzne}, in~\cite{noh_2020} it was found that a noisy quantum circuit initially tends to build up entanglement before reaching a peak, after which the entanglement drops off as the noise takes over. This peak is an example of the `entanglement barrier' observed in the time evolution of noisy quantum systems~\cite{reid2021entanglement,wellnitz2022rise,vovk2024quantum}. 

The metric used by the authors of~\cite{noh_2020} to quantify the amount of entanglement is the \emph{MPO entanglement entropy}, defined as follows. For an $n$-qubit VMPO encoding the state $\kett{\rho}$, let $0 < l < n$ be the index of a bond in the VMPO and write $s_i$ for the $i$th Schmidt coefficient in the Schmidt-decomposition of the state $\kett{\rho}$ with respect to the Hilbert spaces spanned by states of the first $l$ and last $n - l$ qubits, respectively. Equivalently, the $s_i$ are the singular values of the matrix formed by contracting the tensors $M^{[1]}, M^{[2]}, \dots, M^{[n]}$, and reshaping the result into a matrix acting from the space spanned by the first $l$ qubits to the last $n-l$~\footnote{Note that computing the singular values $s_i$ does not require constructing the exponentially large matrix (that maps the first $l$ qubit space that of the last $n-l$ qubits) explicitly. Instead, the VMPO can be brought into so-called canonical form~\cite{schollwock2011density} with respect to the bond $l$, and then the $s_i$ are simply the singular values of the matrix formed by contracting $M^{[l]}$ with $M^{[l+1]}$.}. The MPO entanglement entropy at site $l$ is then given by
\[
    \mS_l(\kett{\rho}) = -\sum_{i=1}^D \frac{s_i^2}{\sum_{j=1}^D s_j^2} \cdot \ln\left( \frac{s_i^2}{\sum_{j=1}^D s_j^2} \right)\,,
\]
where $D$ is the bond dimension of the $l$-th bond. (Note that the rank of the reshaped density matrix is at most $D$). 

This measure of entanglement does not match the von Neumann entropy of the full density matrix, nor of the reduced density matrix on either side of the bond, but can be considered the natural measure of entanglement of a vectorised density matrix in VMPO form which captures both classical and quantum correlations between its constituent qubits. 

The maximum MPO entanglement entropy over all bonds in $\kett{\rho}$:
\[
    \mS_{\max}(\kett{\rho}) = \max_{0 < l < n} \mS_{l}(\kett{\rho})\,,
\]
gives an indication of the maximum bond dimension required to represent $\kett{\rho}$ exactly.  When referring to the MPO entanglement entropy without mentioning a specific bond index $l$, we mean the maximum of the MPO entanglement entropy $\mS_{\max}$ over all bonds. Sometimes the MPO entanglement entropy of a specific bond -- usually the middle bond where the entanglement entropy is typically largest -- is used as an indication of the maximum MPO entanglement entropy $\mS_{\max}$ of the state of interest. We refer the reader to~\cite{noh_2020} for a more detailed discussion on the MPO entanglement entropy.

\subsection{Relationship between emulation fidelity and true fidelity}\label{sec:fidelity_true_fidelity}

In this work we use the numerically-obtained emulation fidelity as a heuristic proxy to the true fidelity. It is not guaranteed to always be a good substitute, however in all systems that we studied we found that the emulation fidelity was indeed a good approximation of the true fidelity and that, in practice, actually seems to provide a lower bound to the true fidelity. We make no theoretical claim here: it is merely an observation that we make from the data that we have obtained. 

Figure~\ref{fig:emulation-fid-vs-actual-fid} shows the typical behavior of the emulation fidelity vs the actual fidelity for a 10-qubit TFIM instance. We observe that indeed the true fidelity is higher than the emulation fidelity.

\clearpage
\section{Further details on the implementation of non-zero noise extrapolation}
\subsection{Quantitative method for determining convergence in (emulation) fidelity}\label{sec:criteria-convergence}
To implement the method explained in Section \ref{ssec:implementation}, for an observable $O$ we compute extrapolated expectation values $\overline{\langle O \rangle}_\lambda$ at noise strength $\lambda$, by performing an extrapolation in emulation fidelity over some range of bond dimensions $D_1, ..., D_k$. An important criteria that we make use of to determine whether to include the extrapolated value $\overline{\langle O \rangle}_\lambda$ computed at noise strength $\lambda$ into $\Lambda$ criterion (\ref{ssec:implementation}) \ref{crit:conv}, which instructs us to only include values of $\lambda$ for which the value of $\langle O \rangle_\lambda$ has approximately converged. 

There are several ways to quantify if the extrapolation is converged. In practice, we have found the following definition to work well. We say that $\overline{\langle O \rangle}_\lambda$ has \emph{approximately converged} if:
\begin{equation}
\left| \frac{ \ex{O}_\lambda^{(D_k)}-\ex{O}_\lambda^{(D_{k-1})}}{\mathcal{F}_\lambda^{(D_k)} - \mathcal{F}_\lambda^{(D_{k-1})}} \right| \cdot (1 - \mathcal{F}_\lambda^{(D_k)}) < \frac{|\ex{O}_\lambda^{(D_k)}|}{2} \, ,
\end{equation}
where, $\ex{O}_\lambda^{(D_j)}$ denotes the expectation value computed at noise strength $\lambda$ and bond dimension $D_j$, and $\mathcal{F}_\lambda^{(D_j)}$ denotes the corresponding fidelity obtained for this emulation.

To see that the above gives a reasonable criteria for ``approximate convergence", note that the first term on the left $\left| \frac{ \ex{O}_\lambda^{(D_k)}-\ex{O}_\lambda^{(D_{k-1})}}{\mathcal{F}_\lambda^{(D_k)} - \mathcal{F}_\lambda^{(D_{k-1})}} \right|  $ gives the magnitude of slope of the straight line fit between the two values that have been computed at the two largest bond dimensions used. This criterion therefore checks that the result obtained from the straight-line fit on the last two data points gives a value that is no more than $\frac{|\ex{O}_\lambda^{(D_k)}|}{2}$ away from the value of $\ex{O}_\lambda^{D_k}$ itself. 

\begin{figure*}[htb!]
    \centering
    \includegraphics[width=0.75\linewidth]{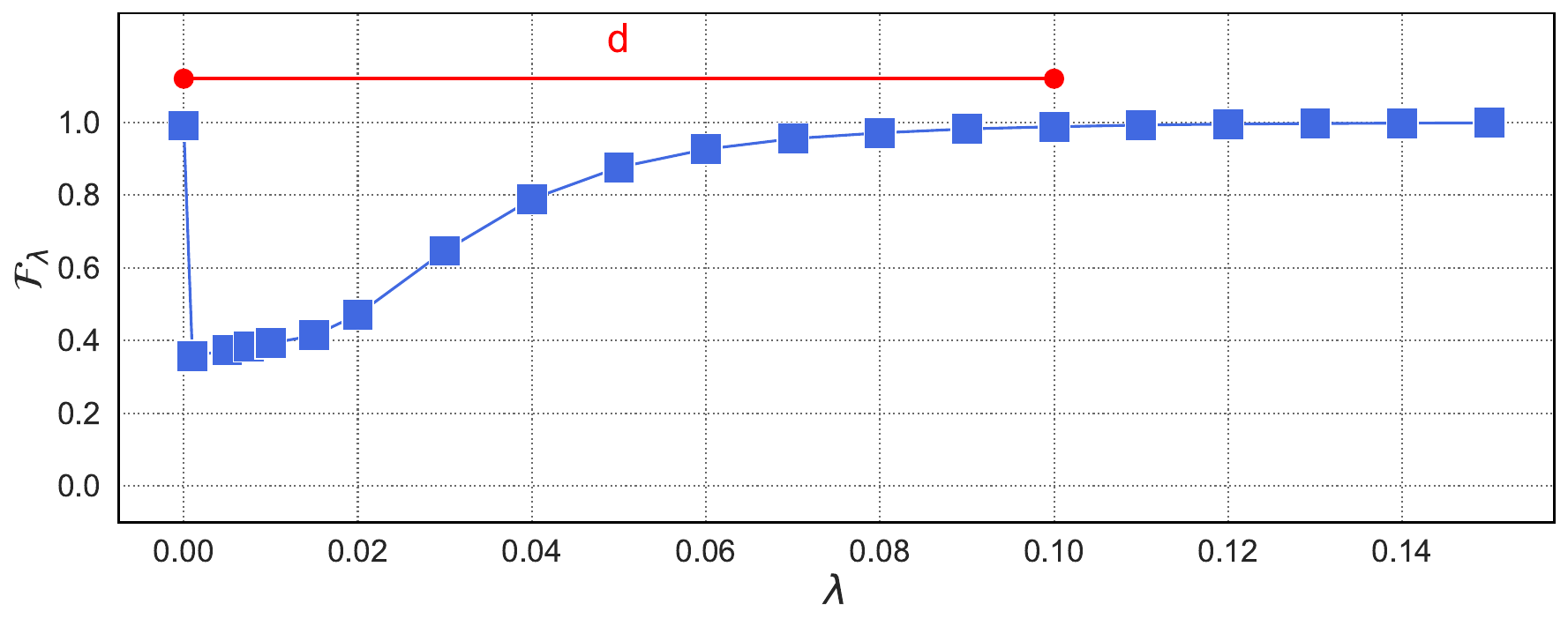}
    \caption{Emulation fidelity $\mF_\lambda$ vs depolarising noise strength $\lambda$. The `dip size' $d$ indicates the regime where $\mathcal{F}_\lambda < 0.99$. Results obtained with $D=32$ for emulations of a circuit implementing 10 Trotter steps of time evolution of a $2 \times 7$ ($14$-qubit) instance of the TFIM (see Appendix~\ref{app:tfim}).}
    \label{fig:fid_vs_noise_strength}
\end{figure*}

\subsection{Further details on the loss function for a log-linear fit}\label{ssec:loss-function-details}
As explained in Section~\ref{ssec:implementation}, the primary method we use to fit the data points to the exponential Ansatz~\eqref{eqn:exp_Ansatz} is to set $c=0$ so that $\hat{O}(\lambda) = a\rm e^{- b\lambda} $ and then fit the logs of the absolute values of the computed observables to a straight line~\footnote{We choose the sign of $a$ to match the sign of $\overline{\langle O\rangle}_{\lambda=0}$ since the target noise strength is always close to the pure state value at $\lambda=0$. When the sign is negative, we use the fitting procedure to determine $|a|$ and then put in the correct sign by hand.}. To perform the fit we need to specify the choice of loss function. 

The most obvious choice of loss function would be the mean-squared error which, we remind the reader, is given by the following equation:
\begin{equation}
    \mathcal{L}_{\rm mse} = \sum_{\lambda \in \Lambda} |- b\lambda + \log(|a|) - \rm log(|\overline{\langle O \rangle}_{\lambda}|)|^2\,.
\end{equation}

This gives equal weighting to the residuals for all values of $\lambda$. However, we find that not all data points carry an equal amount of information. Rather than weighing all datapoints equally, we will instead weight contributions from different values of $\lambda$ differently. To do this we use the weighted mean-square error:
\begin{equation}
    \mathcal{L} = \sum_{\lambda \in \Lambda} w_\lambda^2| - b\lambda + \log(|a|) - \mathrm{log}(|\overline{\langle O \rangle}_{\lambda}|)|^2 \, ,
\end{equation}
where the weights $ w_\lambda$ should have the following properties. To start with (1), we have more confidence in simulations with a higher maximum fidelity $\mathcal{F}_{\lambda}^{\max}$, and therefore the $\omega_{\lambda}$ should increase with increasing $\mathcal{F}_{\lambda}^{\max}$. Second (2), we expect data points closer to the target noise strength $\lambda^*$ to be more relevant than data points far from the target noise strength, and therefore we want the weights $\omega_{\lambda}$ to increase when $|\lambda^* - \lambda|$ decreases. 

To satisfy point (1) above, we choose $\omega_{\lambda}$ to contain a factor of $\mathcal{F}_{\lambda}^{\max}$, to the power of a tunable constant, which we call $\delta_f$. To address point (2), we make the following observation. If we plot the maximally obtained fidelity $\mathcal{F}_{\lambda}^{\max}$ as a function of noise strength $\lambda$, for all systems and noise models emulated for this paper we always observe the same pattern~\footnote{This observed pattern is in agreement with Fig.~\ref{fig:accuracy_v_lambda} that shows the entanglement entropy being low for both the (shallow-circuit) pure and the very noisy simulations, and higher in the more challenging low-noise regime.} -- displayed in Figure~\ref{fig:fid_vs_noise_strength}: we have high maximum fidelity at the pure state point $\lambda = 0$, then an immediate drop, after which the maximum fidelity increases monotonically with increasing noise strength. The observed pattern suggests the presence of an intrinsic \emph{length scale}, which we call the \emph{dip size} $d$; it is defined as the distance between $\lambda = 0$ and the lowest non-zero $\lambda$  such that $\mathcal{F}_\lambda > 0.99$. For example, in Fig.~\ref{fig:fid_vs_noise_strength}, which plots $\mathcal{F}_{\lambda}^{\max}$ against $\lambda$ for a circuit comprised of 10 Trotter steps ($dt = 0.1$) of the Transverse field Ising Hamiltonian on 14 qubits with a depolarizing noise model, we compute the dip size (in red) to be $d = 0.1$. Given this observed length scale, we have found that a natural way to incorporate point (2) above is to let the weights $\omega_{\lambda}$ include a factor that decays exponentially in $|\lambda^* - \lambda| / d$, also to the power of a tunable constant $\delta_d$.

Combining the two points above, we arrive at the following expression for the weights $w_\lambda$ that we use in the loss function:
\begin{equation}
    w_\lambda = (\mathcal{F}_\lambda)^{\delta_f} \cdot e^{-\delta_d \frac{|\lambda^* - \lambda|}{d}}\, ,
\end{equation}
where $\delta_f, \delta_d$ are two tuneable parameters. Note that varying the relative values of $\delta_f$ and $\delta_d$ allows one to emphasise the relative importance of (maximum) emulation fidelity or closeness to target noise strength respectively. For all the results that have been presented in this paper we have used $\delta_f = 2$ and $\delta_d = 20$. These are arbitrary choices and no attempt has been made to optimise these values.

\section{Circuits and noise models}\label{app:circuits-noise-models}

\subsection{Benchmark circuits}

\subsubsection{Transverse field Ising model (TFIM)}\label{app:tfim}
The transverse field Ising model is the ‘quantum version' of the well-known Ising model, and consists of a lattice of interacting spins in the presence of an external (transverse) magnetic field. We focus on the two dimensional case with periodic boundary conditions. The Hamiltonian for the system is
\begin{equation}
        H_{\text{TFIM}} = J \sum_{\langle i,j \rangle} Z_iZ_j + h \sum_{j} X_j
\end{equation}
where $\langle i, j \rangle$ ranges over nearest neighbours on the (periodic) lattice, and $Z, X$ refer to the Pauli operators.

The circuit that we consider implements time-evolution of the model using a second-order Trotter decomposition with $dt=0.25$. The settings that we chose for the evolution were: $J=1$, $h=2$. In all cases the starting state was the all-zeros computational basis state.

\subsubsection{Fermi-Hubbard model (FHM)}\label{app:fhm}
The Fermi-Hubbard model (often just called the Hubbard model in the condensed-matter community) is a simplified model of fermions hopping on a lattice. This model aims to capture the strongly-correlated behaviour of electrons moving between orbitals, and is widely used in the study of phenomena such as high-temperature superconductivity and quantum magnetism. The Hamiltonian consists of a kinetic term allowing for tunneling/`hopping' of particles between lattice sites and a potential term corresponding to an on-site interaction between fermions of different spins. Written in terms of the fermionic ladder operators $a$ and $\ad$, the Hamiltonian is 
\begin{equation}
    \hat{H}_{\text{FHM}} = - t \sum _{\langle i,j \rangle,\sigma }\left( \ad_{i,\sigma } a_{j,\sigma } + \ad_{j,\sigma } a_{i,\sigma }\right) + U \sum _{i} {\hat {n}}_{i\uparrow }{\hat {n}}_{i\downarrow }
\end{equation}
where $\langle i, j \rangle$ ranges over neighbouring sites on the lattice, $\sigma \in \{\uparrow, \downarrow\}$ specifies the spin, and the number operators $\hat{n}_{i\sigma} = \ad_{i\sigma}a_{i\sigma}$,  count the number (0 or 1) of fermions occupying the $\sigma$-orbital on the $i$th lattice site. 

To simulate the Fermi-Hubbard model on quantum computers, one needs to map the fermionic Hamiltonian to a qubit one. There are a number of fermion-to-qubit mappings, with the simplest being the well-known Jordan-Wigner (JW) encoding. In this work we consider the Fermi-Hubbard Hamiltonian under the JW encoding, in which one qubit is assigned to store the occupation of a single fermionic mode (choice of lattice site $i$ and spin $\sigma$). For an $m \times n$ lattice, let the (bijective) map between fermioniq modes $(i, \sigma)$ and qubit indices $k$ be $\mathcal{J} : \{0, \dots, n-1\} \times \{0, \dots, m-1\} \times \{\uparrow, \downarrow\} \mapsto \{0, \dots, 2nm - 1\}$. Then we can write the Hubbard Hamiltonian under the JW encoding as:
\begin{equation}
    H_{\text{FHM}} = - \frac{t}{2} \sum _{\langle i, j \rangle, \sigma} (X_{\mathcal{J}(i,\sigma)}X_{\mathcal{J}(j,\sigma)} + Y_{\mathcal{J}(i,\sigma)}Y_{\mathcal{J}(j,\sigma)})\overrightarrow{Z}_{\mathcal{J}(i,\sigma),\mathcal{J}(j,\sigma)} + \frac{U}{4} \sum_{i} (I_{\mathcal{J}(i,\uparrow)} - Z_{\mathcal{J}(i,\uparrow)})(I_{\mathcal{J}(i,\downarrow)} - Z_{\mathcal{J}(i,\downarrow)})
\end{equation}
where I, X, Y and Z are the Pauli operators with their subscripts denoting the site they act on, and
\begin{equation}
    \overrightarrow{Z}_{\mathcal{J}(i,\sigma),\mathcal{J}(j,\sigma)} = \prod_{k = k_{\min}(i,j,\sigma) + 1}^{k_{\max}(i,j,\sigma) - 1} Z_k
\end{equation}
with $k_{\min}(i, j, \sigma) = \min(\mathcal{J}(i, \sigma), \mathcal{J}(j,\sigma))$ and $k_{\max}(i, j, \sigma) = \max(\mathcal{J}(i, \sigma), \mathcal{J}(j,\sigma))$, is a string of Pauli-$Z$ operators acting on all qubits between $\mathcal{J}(i,\sigma)$ and $\mathcal{J}(j,\sigma)$. 

In this paper we consider the Fermi-Hubbard model with open boundary conditions on a 2-dimensional lattice with $m\times n$ sites at half-filling (i.e.~in the subspace spanned by particle number $mn$, or equivalently in the subspace spanned by Hamming-weight $mn$ computational basis states). The circuits that we consider implement time evolution of the model via first-order Trotterization, implemented on a Google Sycamore-like~\cite{arute2019quantum} 2D nearest-neighbour qubit architecture. In an effort to keep the depth of the circuits low, we employ fermionic swap networks as in~\cite{cai2020resource,cade2020}, which allows one to implement the vertical hopping terms without explicitly including the strings of Pauli-$Z$ operators. For all circuits, the Trotter step size was set to $0.1$ and the starting state was a (classical) anti-ferromagnetic product state at half filling with neighbouring lattice sites being occupied by fermions of opposite spins, i.e.
\[
    \begin{pmatrix}
        \uparrow & \downarrow & \uparrow & \cdots \\
        \downarrow & \uparrow & \downarrow & \cdots \\
        \uparrow & \downarrow & \uparrow & \cdots \\
        \vdots & \vdots & \vdots & \ddots \\
    \end{pmatrix} \,.
\]

\subsubsection{$XY$ model (XYM)}\label{app:xym}
The (isotropic) $XY$ model has the Hamiltonian
\begin{equation}
        H_{\text{XYM}} = J \sum_{\langle i, j \rangle} (X_iX_j + Y_iY_j) + h \sum_i Z_i
\end{equation}
where $\langle i, j \rangle$ ranges over neighbouring sites on the lattice.
In one dimension, the model is known to be exactly solvable via a mapping to a system of non-interacting fermions. This means that its time evolution can be simulated efficiently classically. Indeed, it is straightforward to check that the quantum circuit implementing its (e.g.~second-order trotterised) time evolution is an instance of the family of so-called matchgate circuits (see Appendix~\ref{app:matchgate} below), which are classically simulable in polynomial time in the number of qubits, again via a mapping to a system of non-interacting fermions. In particular, this allows us to classically compute expectation values of Pauli observables for large system sizes, a fact that we exploit in Section~\ref{sec:results_large_xy} to evaluate the accuracy of our zero-noise extrapolation method for circuits of up to 60 qubits. 

The circuits that we consider in this work implement time evolution of the $XY$ model with $J=0.5$ and $h=0.23$ via a second-order Trotter decomposition with step size $dt=0.1$ and $30$ Trotter steps. The initial state we use is the anti-ferromagnetically ordered product state which is a computational basis state $\ket{1010\dots}$ with alternating 1s and 0s on the qubits.

\subsection{Noise models}
\subsubsection{Depolarizing noise model}\label{app:depolarizing}
Consider the noise model $\ndp$ characterised by the application of two-qubit depolarizing noise channels with parameter $\lambda$ applied after every two-qubit gate in the circuit (one-qubit gates have no channel applied before/after them). Such a model is common in the quantum computing literature, and often captures many of the properties of noisy hardware, in which one-qubit gates can be implemented with relatively high fidelity whereas two-qubit gates introduce the majority of the unwanted noise effects~\footnote{Although note that in most hardware idling also proves to be a significant source of error, which is not included in this noise model.}. The noise channel used in this noise model has Kraus operators
\begin{equation}
      K_0 := \sqrt{1 - \frac{15\lambda}{16}} I \qquad K_{ij} = \sqrt{\frac{\lambda}{16}} \sigma_i \sigma_j
\end{equation}
where $\sigma_k$ is the $k$th Pauli operator: $\sigma_0 = I$, $\sigma_1 = X$, $\sigma_2 = Y$, $\sigma_3 = Z$ and $i, j$ each range over $0,1,2,3$ subject to $i + j \neq 0$. This channel acts on a two-qubit state $\rho$ as 
\begin{equation}
    \rho \mapsto (1 - \lambda) \rho + \lambda \frac{I}{4}
\end{equation}

\subsubsection{Cat-qubit noise model}\label{app:cat_noise}
Quantum hardware based on cat-qubits experience a biased noise model dominated by dephasing errors, but are naturally protected from bit-flip errors~\cite{guillaud_2021,guillaud2023quantum}. We consider a noise model $\ncat$ based on one given by~\cite{guillaud_2021,bornens2023variational}, which is summarised in Table~\ref{tab:cat_noise}.

\begin{table*}[t]
    \centering
        \begin{tabular}{ |p{1cm}||p{3.5cm}|p{3.5cm}|p{3.5cm}|  }
         \hline
         \multicolumn{4}{|c|}{Errors per gate for noise strength $\lambda$} \\
         \hline
         Gate & Z($\theta$) & H & CX\\
         \hline
           & Error \hspace{0.4cm} Probability    &Error \hspace{0.4cm} Probability&   Error \hspace{0.4cm} Probability\\
         &   I \hspace{1.7cm} $1 - \lambda$  & I \hspace{1.7 cm} $1 - 5\lambda$  &I \hspace{1.7 cm} $1 - 4\lambda$ \\
          & Z \hspace{1.7cm} $\lambda$  & Z \hspace{1.7cm} $3\lambda$ & $Z_1$ \hspace{1.5 cm} $3\lambda$ \\
            & & X \hspace{1.7cm} $2\lambda$ & $Z_2$ \hspace{1.5 cm} $\lambda/2$ \\
         &    & & $Z_1 Z_2$ \hspace{1.1 cm} $\lambda/2$\\
         \hline
        \end{tabular}
        \caption{Noise model $\ncat$ for cat-qubits including Hadamard gate. Based on Table~1 of ref.~\cite{bornens2023variational}.}
\label{tab:cat_noise}
\end{table*}

\subsection{Matchgate circuits, trajectory-based emulation, and matchgate noise models}\label{app:matchgate}
Matchgate circuits are a family of quantum circuits known to be classically simulable, under certain conditions. A matchgate circuit is a quantum circuit consisting of only matchgates acting on neighbouring qubits on a line. A matchgate $G(A,B)$ is of the form
\begin{equation}
    G(A, B) = \begin{pmatrix}
                    p & 0 & 0 & q \\
                    0 & w & x & 0 \\
                    0 & y & z & 0 \\
                    r & 0 & 0 & s \\
                \end{pmatrix}
\end{equation}
where
\begin{equation}
    A = \begin{pmatrix}
            p & q \\
            r & s \\
        \end{pmatrix}
\end{equation}
and
\begin{equation}
    B = \begin{pmatrix}
            w & x \\
            y & z \\
        \end{pmatrix}
\end{equation}
are unitaries that satisfy $\det(A) = \det(B)$. All matchgates preserve the odd- and even-parity subspaces of the qubits that they act upon. Any quantum computation where: 
\begin{itemize}
    \item The initial state is a computational basis state;
    \item All gates are nearest neighbour matchgates (in 1D);
    \item The output is a single- or two-qubit Pauli measurement;
\end{itemize}
can be simulated efficiently classically~\cite{jozsa2008matchgates}. The conditions on the input and output can be loosened somewhat whilst retaining the result, but for our purposes this basic result is enough.

To perform the noisy emulations, we make use of a trajectory-sampling method: each time a gate is applied, we sample over Kraus operators from the noise channels and stochastically insert them into the circuit in the relevant place. Concretely, for a noise channel $C$ with Kraus operators $K_0,\dots,K_k$ the sampling is performed according to the probabilities $p_0,\dots,p_k$ defined as
\begin{equation}
    p_i = \braket{\psi | K_i^\dagger K_i | \psi}
\end{equation}

for $i \in \{0, \ldots, k\}$, where $\ket{\psi}$ is the state of the computation at the point that the noise channel $C$ appears in the circuit. When the $K_i$ are scalar multiples of unitary operators (as they always are for the cases we consider), this inner product reduces to $p_i = K_i^\dagger K_i$, and we can sample the Kraus operators ahead of time before performing the emulation. We refer the reader to ref.~\cite{isakov2021simulations} for a more detailed explanation of noisy emulation via the trajectory method. 

For each sampled circuit (trajectory), one can compute the exact expectation value of one or more Pauli observables. By averaging over these values, we approximate the expectation value of the observable with respect to the state which would be obtained by applying the noisy circuit to the same initial state. Note that it can often require very many samples to reach convergence and hence obtain an accurate estimate of the expectation value. For instance, we found for one particular 60-qubit circuit of depth 121 and 1770 two-qubit gates, around $10^5$ samples were required to reach convergence for a two-qubit Pauli observable.

\subsubsection{Matchgate noise}\label{app:matchgate-noise}
In Section~\ref{sec:results}, we want to use the classical simulability of matchgate circuits as a method for evaluating the accuracy of non-zero noise extrapolation for estimating expectation values from large noisy quantum circuits. This requires us to identify an appropriate noise model that can be applied to a matchgate circuit while preserving its classical simulability. 

The usual depolarizing noise model (Appendix~\ref{app:depolarizing}) will be our starting point. In its basic form it cannot be used, since the Kraus operators $IX, IY, ZX, ZY$ and their Hermitian conjugates do not correspond to valid matchgates. However, one can simply remove these Kraus operators to obtain a `matchgate depolarizing channel'. Our matchgate depolarizing noise model $\mathcal{N}_{\text{mg-dep}}(\lambda)$ corresponds to the application of `matchgate depolarizing' channels after every two-qubit gate with strength $\lambda$. The channel as defined via its Kraus operators is:
\begin{equation}
    K_0 = \sqrt{1 - \lambda} I \qquad K_{\sigma_k} = \sqrt{\frac{\lambda}{7}} \sigma_k
\end{equation}
where $\sigma_k \in \{Z \otimes I, I \otimes Z, X \otimes X, X \otimes Y, Y \otimes Y, Y \otimes X, Z \otimes Z\}$. It is easy to see that this channel is unital, and therefore we  would expect the model to affect values of local observables in a similar way to the depolarizing and cat-qubit noise models.

\begin{figure*}[htb!]
    \centering
    \subfigure{\label{fig:a}\includegraphics[width=0.5\textwidth]{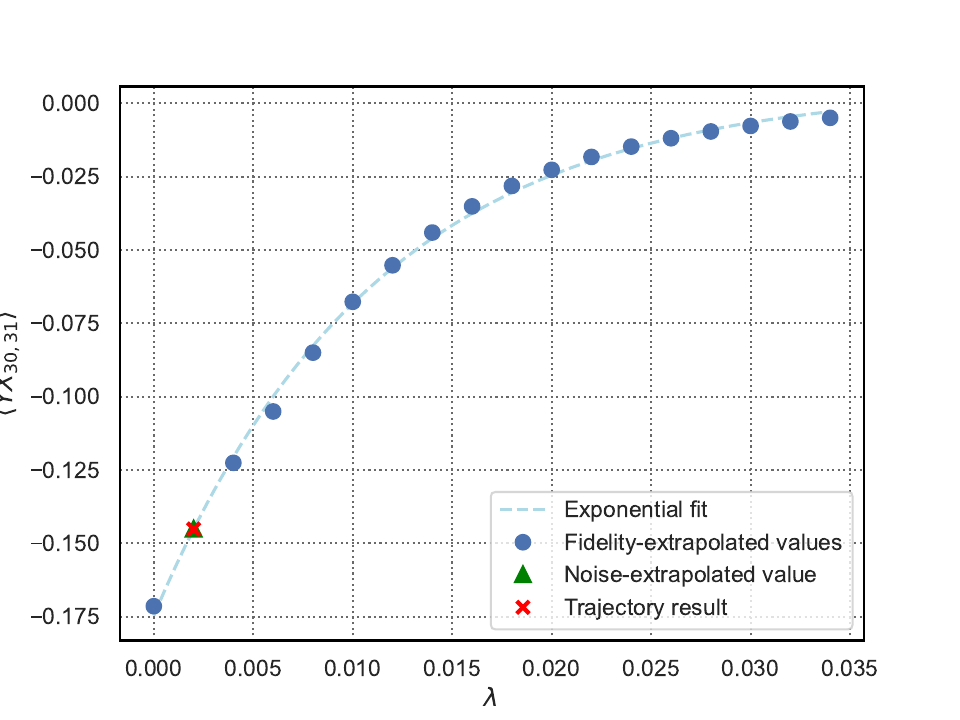}}%
    ~
    \subfigure{\label{fig:b}\includegraphics[width=0.5\textwidth]{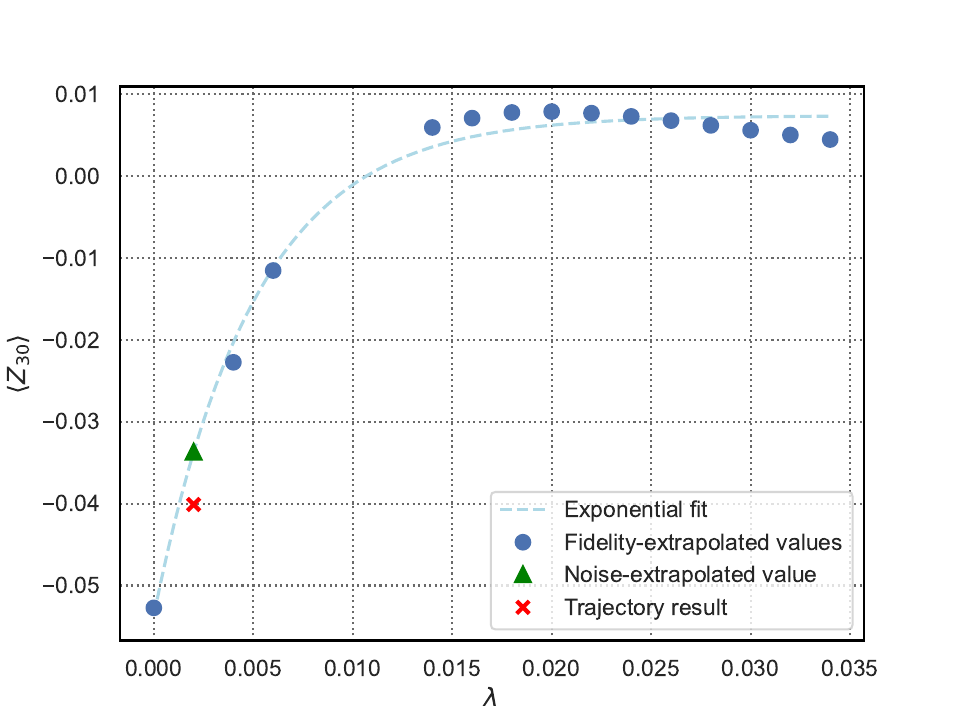}}
    \caption{Value of (a) $\ex{YX_{30, 31}}$ and (b) $\ex{Z_{30}}$ obtained via fidelity extrapolation for different noise strengths $\lambda$, with a fit to the Ansatz $f(\lambda) = a e^{-b \lambda} + c$ (dashed blue lines) resulting in noise-extrapolated estimates of $\ex{YX_{30, 31}}$ and $\ex{Z_{30}}$ at $\lambda^*=0.002$ (green triangles), and the values of $\ex{YX_{30, 31}}$ and $\ex{Z_{30}}$ at $\lambda^*=0.002$ obtained via trajectory simulations (red crosses).}
    \label{fig:XYM_z_expval_exp}
\end{figure*}

\section{Further details on the results}\label{app:further_results}
\subsection{$XY$ model}\label{ssec:xym_results}
In Section~\ref{sec:results_large_xy} we presented the results of performing a non-zero noise extrapolation by using a weighted fit of a straight line to a log-plot of the expectation values. For the observable $\ex{Z}_{30}$ in particular, this method of fitting meant that we had to rely on only a few data points where the sign of the estimated expectation values matched the $\lambda = 0$ one. Despite obtaining a reliable estimate of the expectation value in this way (within 2 standard deviations of the result obtained via a trajectory simulation), it is also possible to fit directly to the exponential Ansatz~\eqref{eqn:exp_Ansatz} , as mentioned in Section~\ref{ssec:implementation}. In Figures~\ref{fig:XYM_z_expval_exp} (a) \& (b) we show the fidelity-extrapolated expectation values of $\ex{YX_{30, 31}}$ and $\ex{Z_{30}}$ for each value of $\lambda$ (again removing the anomalous $\lambda^* = 0.002$ point and others according to criterion \ref{crit:conv} from Section~\ref{ssec:implementation}), with an unweighted least-squares fit to the Ansatz $f(\lambda) = a e^{-b \lambda} + c$, where $a, b, c$ are the parameters to be fit. The extrapolated expectation values obtained using this method for $\lambda^* = 0.002$ were $-0.1452$ and $-0.0348$, respectively, close to those obtained via the fitting method described in the main text.

\subsection{Fermi-Hubbard model}
\label{app:further_hubbard_results}
In Figure~\ref{fig:large_quiver} of Section~\ref{ssec:fhm_large_results} we visualised data from simulations of (noisy) circuits implementing time evolution of the Fermi-Hubbard model. Tables~\ref{tab:small_hubbard_filling}-\ref{tab:small_hubbard_magnetization} below contain the data used to generate the visualizations for the $2 \times 4$ instance of the FHM (Figure~\ref{fig:small_quiver}). Tables~\ref{tab:hubbard_large_nzne_fillings}-\ref{tab:hubbard_large_nzne_magnetizations} contain the data used to generate the visualizations for the $5 \times 6$ instance of the FHM when non-zero noise extrapolation was used (Figure~\ref{fig:large_quiver}). Tables~\ref{tab:hubbard_large_single_emulation_fillings}-\ref{tab:hubbard_large_single_emulation_magnetizations} contain the data used to generate the visualizations for the $5 \times 6$ instance of the FHM when only a single emulation was used (including the noiseless emulation) (Figure~\ref{fig:large_quiver}). 

Finally, we make a clarification on our extrapolation method to generate this data. As we saw in Section~\ref{ssec:xym_results}, when the value of an expectation value at the target noise strength is close to zero, it can happen that many data points cannot be used for the noise extrapolation because the estimated values of expectation values at surrounding noise strengths often come with different signs (see e.g. Figure~\ref{fig:XYM_z_expval_exp} for an illustration of this). In extreme cases, the noise extrapolation approach that we presented in the main text does not work at all, in particular when only one or zero data points remain after applying criterion~\ref{crit:sign}. For such cases, as explained in Section~\ref{ssec:implementation}, we fall back to the method of fitting directly to the exponential Ansatz~\eqref{eqn:exp_Ansatz}, using the unweighted mean-square error as the loss function. To generate the data in this section, we use whenever possible the log-linear fitting and extrapolation approach given in Section~\ref{ssec:implementation}. When this fails, we use the alternative method of fitting directly to an exponential.

\clearpage

\begin{table}[htb]
    \centering

    \begin{minipage}{0.45\textwidth}
        \centering
        \begin{subtable}{}
        
        \caption*{$\lambda = 0.0$}
        \label{tab:0.0}
        \begin{tabular}{lrr}
        \toprule
         & 0 & 1 \\
        \midrule
        0 & 1.000065 & 0.999936 \\
        1 & 1.000564 & 0.999446 \\
        2 & 0.992713 & 1.007330 \\
        3 & 1.005346 & 0.994599 \\
        \bottomrule
        \end{tabular}
        
        \end{subtable}%
        \begin{subtable}{}
        
        \caption*{$\lambda = 0.001$}
        \label{tab:0.001}
        \begin{tabular}{lrr}
        \toprule
         & 0 & 1 \\
        \midrule
        0 & 1.003067 & 0.997421 \\
        1 & 1.000806 & 0.998741 \\
        2 & 0.992681 & 1.008126 \\
        3 & 1.007442 & 0.991519 \\
        \bottomrule
        \end{tabular}
        
        \end{subtable}%
        \begin{subtable}{}
        
        \caption*{$\lambda = 0.005$}
        \label{tab:0.005}
        \begin{tabular}{lrr}
        \toprule
         & 0 & 1 \\
        \midrule
        0 & 1.008854 & 0.993742 \\
        1 & 0.999375 & 0.998942 \\
        2 & 0.995535 & 1.006475 \\
        3 & 1.008463 & 0.988227 \\
        \bottomrule
        \end{tabular}
        
        \end{subtable}%
        \begin{subtable}{}
        
        \caption*{$\lambda = 0.01$}
        \label{tab:0.01}
        \begin{tabular}{lrr}
        \toprule
         & 0 & 1 \\
        \midrule
        0 & 1.007586 & 0.996158 \\
        1 & 0.998897 & 0.998751 \\
        2 & 0.999213 & 1.003625 \\
        3 & 1.004473 & 0.991223 \\
        \bottomrule
        \end{tabular}
        
        \end{subtable}%
        \caption{\textbf{Filling}}
        \label{tab:small_hubbard_filling}
    \end{minipage}
    \hfill
    \begin{minipage}{0.45\textwidth}
        \centering
        \begin{subtable}{}
        \caption*{$\lambda = 0.0$}
        \label{tab:0.0}
        \begin{tabular}{lrr}
        \toprule
         & 0 & 1 \\
        \midrule
        0 & 0.284189 & -0.284158 \\
        1 & -0.216263 & 0.216275 \\
        2 & 0.248565 & -0.248554 \\
        3 & -0.166862 & 0.166937 \\
        \bottomrule
        \end{tabular}
        
        \end{subtable}%
        \begin{subtable}{}
        
        \caption*{$\lambda = 0.001$}
        \label{tab:0.001}
        \begin{tabular}{lrr}
        \toprule
         & 0 & 1 \\
        \midrule
        0 & 0.236285 & -0.236588 \\
        1 & -0.166204 & 0.166370 \\
        2 & 0.196034 & -0.196506 \\
        3 & -0.139009 & 0.138994 \\
        \bottomrule
        \end{tabular}
        
        \end{subtable}%
        \begin{subtable}{}
        
        \caption*{$\lambda = 0.005$}
        \label{tab:0.005}
        \begin{tabular}{lrr}
        \toprule
         & 0 & 1 \\
        \midrule
        0 & 0.115632 & -0.115538 \\
        1 & -0.057773 & 0.057787 \\
        2 & 0.076451 & -0.076839 \\
        3 & -0.069819 & 0.069996 \\
        \bottomrule
        \end{tabular}
        
        \end{subtable}%
        \begin{subtable}{}
        
        \caption*{$\lambda = 0.01$}
        \label{tab:0.01}
        \begin{tabular}{lrr}
        \toprule
         & 0 & 1 \\
        \midrule
        0 & 0.047395 & -0.046952 \\
        1 & -0.014927 & 0.014952 \\
        2 & 0.022936 & -0.023452 \\
        3 & -0.028216 & 0.028818 \\
        \bottomrule
        \end{tabular}
        \end{subtable}%
        \caption{\textbf{Magnetization}}
        \label{tab:small_hubbard_magnetization}
    \end{minipage}%

    \label{tab:side_by_side}
\end{table}


\begin{table}
\centering 
    \begin{subtable}{}

    \caption*{$\lambda = 0.001$}
    \label{tab:hubbard_large_nzne_fillings_0.001}
    \begin{tabular}{lrrrrrr}
    \toprule
     & 0 & 1 & 2 & 3 & 4 & 5 \\
    \midrule
    0 & 1.008353 & 1.027768 & 0.957620 & 0.974604 & 1.027919 & 1.007470 \\
    1 & 0.999917 & 1.015517 & 0.968009 & 0.969184 & 1.015030 & 1.011040 \\
    2 & 1.005741 & 1.024243 & 0.997083 & 0.961882 & 1.024213 & 1.005428 \\
    3 & 1.009583 & 1.028761 & 0.969116 & 0.967932 & 1.027398 & 1.010168 \\
    4 & 1.013536 & 1.027895 & 0.973385 & 0.972292 & 1.027792 & 1.008363 \\
    \bottomrule
    \end{tabular}
    
    \end{subtable}%
    \begin{subtable}{}
    
    \caption*{$\lambda = 0.005$}
    \label{tab:hubbard_large_nzne_fillings_0.005}
    \begin{tabular}{lrrrrrr}
    \toprule
     & 0 & 1 & 2 & 3 & 4 & 5 \\
    \midrule
    0 & 1.001110 & 1.005230 & 0.985888 & 0.992401 & 1.005710 & 1.000917 \\
    1 & 1.006382 & 1.003088 & 0.994141 & 0.995142 & 1.011379 & 1.005853 \\
    2 & 0.999770 & 1.004337 & 1.000001 & 0.985771 & 1.004302 & 0.999085 \\
    3 & 1.003338 & 1.009945 & 0.995057 & 0.994047 & 1.007851 & 1.003971 \\
    4 & 0.999357 & 1.005685 & 0.986461 & 0.990159 & 1.005247 & 1.001114 \\
    \bottomrule
    \end{tabular}
    
    \end{subtable}%
    \begin{subtable}{}
    
    \caption*{$\lambda = 0.01$}
    \label{tab:hubbard_large_nzne_fillings_0.01}
    \begin{tabular}{lrrrrrr}
    \toprule
     & 0 & 1 & 2 & 3 & 4 & 5 \\
    \midrule
    0 & 0.999978 & 1.000720 & 0.996430 & 0.998087 & 1.000830 & 0.999968 \\
    1 & 1.007230 & 1.001784 & 0.999293 & 0.999514 & 1.009513 & 1.002137 \\
    2 & 0.999562 & 1.000541 & 1.000002 & 0.995799 & 1.000536 & 0.999158 \\
    3 & 1.000823 & 1.002645 & 0.999493 & 0.999269 & 1.001646 & 1.001001 \\
    4 & 0.996264 & 1.000823 & 0.990899 & 0.996896 & 1.000724 & 0.999979 \\
    \bottomrule
    \end{tabular}
    
    \end{subtable}%

    \caption{\textbf{Filling} Non-zero noise extrapolation}
    \label{tab:hubbard_large_nzne_fillings}
\end{table}
    
\begin{table}
\centering 
    \begin{subtable}{}
    \caption*{$\lambda = 0.001$}
    \label{tab:hubbard_large_nzne_magnetizations_0.001}
    \begin{tabular}{lrrrrrr}
    \toprule
     & 0 & 1 & 2 & 3 & 4 & 5 \\
    \midrule
    0 & -0.029654 & 0.016451 & -0.004851 & 0.020404 & -0.017759 & 0.029102 \\
    1 & 0.016346 & -0.005690 & 0.003341 & -0.004452 & 0.004702 & -0.028542 \\
    2 & -0.025573 & 0.014605 & 0.001830 & 0.005275 & -0.014585 & 0.025996 \\
    3 & 0.027154 & -0.018816 & 0.004726 & -0.003297 & 0.017694 & -0.026569 \\
    4 & -0.022994 & 0.017755 & 0.008883 & 0.021820 & -0.016452 & 0.029689 \\
    \bottomrule
    \end{tabular}
    
    \end{subtable}%
    \begin{subtable}{}
    
    \caption*{$\lambda = 0.005$}
    \label{tab:hubbard_large_nzne_magnetizations_0.005}
    \begin{tabular}{lrrrrrr}
    \toprule
     & 0 & 1 & 2 & 3 & 4 & 5 \\
    \midrule
    0 & -0.007298 & 0.001940 & -0.001435 & 0.007600 & -0.002756 & 0.006904 \\
    1 & 0.009218 & -0.000115 & 0.000203 & -0.001163 & 0.007545 & -0.009739 \\
    2 & -0.006397 & 0.001837 & 0.000001 & 0.003273 & -0.001778 & 0.007167 \\
    3 & 0.007321 & -0.006859 & 0.001345 & -0.000149 & 0.005014 & -0.006750 \\
    4 & -0.008434 & 0.002746 & -0.003476 & 0.009843 & -0.001948 & 0.007311 \\
    \bottomrule
    \end{tabular}
    
    \end{subtable}%
    \begin{subtable}{}
    
    \caption*{$\lambda = 0.01$}
    \label{tab:hubbard_large_nzne_magnetizations_0.01}
    \begin{tabular}{lrrrrrr}
    \toprule
     & 0 & 1 & 2 & 3 & 4 & 5 \\
    \midrule
    0 & -0.001298 & 0.000018 & -0.000306 & 0.001915 & -0.000199 & 0.001170 \\
    1 & 0.007541 & -0.001117 & -0.000037 & -0.000175 & 0.008403 & -0.002729 \\
    2 & -0.001247 & 0.000078 & 0.000000 & 0.001424 & -0.000063 & 0.001672 \\
    3 & 0.001446 & -0.001931 & 0.000212 & 0.000053 & 0.001036 & -0.001298 \\
    4 & -0.004864 & 0.000198 & -0.007019 & 0.003107 & -0.000019 & 0.001301 \\
    \bottomrule
    \end{tabular}
    
    \end{subtable}%
    \caption{\textbf{Magnetization} Non-zero noise extrapolation.}
    \label{tab:hubbard_large_nzne_magnetizations}
\end{table}


\begin{table}
\centering 
    \begin{subtable}{}

        \caption*{$\lambda = 0.0$}
        \label{tab:hubbard_large_single_emulation_fillings_0.0}
        \begin{tabular}{lrrrrrr}
        \toprule
         & 0 & 1 & 2 & 3 & 4 & 5 \\
        \midrule
        0 & 1.013202 & 1.042549 & 0.944208 & 0.945521 & 1.041727 & 1.011977 \\
        1 & 1.012266 & 1.037471 & 0.951062 & 0.951024 & 1.037435 & 1.012233 \\
        2 & 1.010462 & 1.037518 & 0.951211 & 0.951166 & 1.037561 & 1.010537 \\
        3 & 1.012296 & 1.037521 & 0.951057 & 0.951098 & 1.037447 & 1.012211 \\
        4 & 1.011985 & 1.041728 & 0.945522 & 0.944219 & 1.042556 & 1.013214 \\
        \bottomrule
        \end{tabular}
        
        \end{subtable}%
        \begin{subtable}{}
        
        \caption*{$\lambda = 0.001$}
        \label{tab:hubbard_large_single_emulation_fillings_0.001}
        \begin{tabular}{lrrrrrr}
        \toprule
         & 0 & 1 & 2 & 3 & 4 & 5 \\
        \midrule
        0 & 1.001209 & 1.002122 & 0.996437 & 0.996792 & 1.001949 & 1.001451 \\
        1 & 1.001204 & 1.002425 & 0.995388 & 0.995756 & 1.003124 & 1.001703 \\
        2 & 1.001023 & 1.002191 & 0.997083 & 0.997137 & 1.001941 & 1.000694 \\
        3 & 1.001387 & 1.003415 & 0.995087 & 0.995279 & 1.003462 & 1.001268 \\
        4 & 1.001108 & 1.002197 & 0.997083 & 0.997005 & 1.001951 & 1.000712 \\
        \bottomrule
        \end{tabular}
        
        \end{subtable}%
        \begin{subtable}{}
        
        \caption*{$\lambda = 0.005$}
        \label{tab:hubbard_large_single_emulation_fillings_0.005}
        \begin{tabular}{lrrrrrr}
        \toprule
         & 0 & 1 & 2 & 3 & 4 & 5 \\
        \midrule
        0 & 1.011243 & 1.011769 & 1.010229 & 1.010806 & 1.010013 & 1.010195 \\
        1 & 0.997848 & 0.998413 & 0.997091 & 0.997684 & 0.996999 & 0.997088 \\
        2 & 1.009172 & 1.009004 & 1.009505 & 1.009069 & 1.009862 & 1.009643 \\
        3 & 0.998409 & 0.999201 & 0.997276 & 0.998185 & 0.997034 & 0.997246 \\
        4 & 1.009634 & 1.009160 & 1.010504 & 1.009631 & 1.011551 & 1.010980 \\
        \bottomrule
        \end{tabular}
        
        \end{subtable}%
        \begin{subtable}{}
        
        \caption*{$\lambda = 0.01$}
        \label{tab:hubbard_large_single_emulation_fillings_0.01}
        \begin{tabular}{lrrrrrr}
        \toprule
         & 0 & 1 & 2 & 3 & 4 & 5 \\
        \midrule
        0 & 1.000452 & 1.000432 & 1.000473 & 1.000438 & 1.000484 & 1.000463 \\
        1 & 0.999749 & 0.999753 & 0.999748 & 0.999750 & 0.999748 & 0.999746 \\
        2 & 1.000332 & 1.000330 & 1.000332 & 1.000330 & 1.000334 & 1.000334 \\
        3 & 0.999768 & 0.999769 & 0.999775 & 0.999771 & 0.999779 & 0.999774 \\
        4 & 1.000468 & 1.000480 & 1.000460 & 1.000480 & 1.000463 & 1.000476 \\
        \bottomrule
        \end{tabular}
        
        \end{subtable}%

    \caption{\textbf{Filling} Single emulation}
    \label{tab:hubbard_large_single_emulation_fillings}
    \end{table}

\begin{table}
\centering 
    \begin{subtable}{}
    
    \caption*{$\lambda = 0.0$}
    \label{tab:hubbard_large_single_emulation_magnetizations_0.0}
    \begin{tabular}{lrrrrrr}
    \toprule
     & 0 & 1 & 2 & 3 & 4 & 5 \\
    \midrule
    0 & -0.042217 & 0.027136 & -0.006563 & 0.005865 & -0.027891 & 0.041802 \\
    1 & 0.037755 & -0.024221 & 0.005958 & -0.005891 & 0.024204 & -0.037729 \\
    2 & -0.036544 & 0.024012 & -0.005753 & 0.005618 & -0.024106 & 0.036516 \\
    3 & 0.037759 & -0.024194 & 0.005992 & -0.005971 & 0.024250 & -0.037772 \\
    4 & -0.041770 & 0.027908 & -0.005828 & 0.006598 & -0.027116 & 0.042260 \\
    \bottomrule
    \end{tabular}
    
    \end{subtable}%
    \begin{subtable}{}
    
    \caption*{$\lambda = 0.001$}
    \label{tab:hubbard_large_single_emulation_magnetizations_0.001}
    \begin{tabular}{lrrrrrr}
    \toprule
     & 0 & 1 & 2 & 3 & 4 & 5 \\
    \midrule
    0 & -0.003791 & 0.000626 & 0.001843 & -0.001783 & -0.000685 & 0.003868 \\
    1 & 0.004165 & -0.002011 & -0.000618 & 0.000295 & 0.002320 & -0.003992 \\
    2 & -0.004575 & 0.000377 & 0.001830 & -0.001845 & -0.000072 & 0.004169 \\
    3 & 0.003768 & -0.001700 & -0.000631 & 0.000520 & 0.002090 & -0.004047 \\
    4 & -0.003785 & 0.000692 & 0.002608 & -0.002893 & -0.000092 & 0.003651 \\
    \bottomrule
    \end{tabular}
    
    \end{subtable}%
    \begin{subtable}{}
    
    \caption*{$\lambda = 0.005$}
    \label{tab:hubbard_large_single_emulation_magnetizations_0.005}
    \begin{tabular}{lrrrrrr}
    \toprule
     & 0 & 1 & 2 & 3 & 4 & 5 \\
    \midrule
    0 & -0.004258 & -0.010199 & 0.008396 & -0.002758 & 0.015724 & 0.009867 \\
    1 & 0.008001 & 0.013790 & -0.003801 & 0.006616 & -0.010910 & -0.005203 \\
    2 & -0.007003 & -0.013619 & 0.006756 & -0.005366 & 0.015018 & 0.008412 \\
    3 & 0.007503 & 0.014427 & -0.006928 & 0.005797 & -0.015575 & -0.008598 \\
    4 & -0.006477 & -0.012246 & 0.005743 & -0.005048 & 0.012903 & 0.007312 \\
    \bottomrule
    \end{tabular}
    
    \end{subtable}%
    \begin{subtable}{}
    
    \caption*{$\lambda = 0.01$}
    \label{tab:hubbard_large_single_emulation_magnetizations_0.01}
    \begin{tabular}{lrrrrrr}
    \toprule
     & 0 & 1 & 2 & 3 & 4 & 5 \\
    \midrule
    0 & -0.000198 & -0.000308 & 0.000046 & -0.000169 & 0.000190 & 0.000080 \\
    1 & 0.000015 & 0.000059 & -0.000087 & 0.000001 & -0.000145 & -0.000100 \\
    2 & -0.000115 & -0.000192 & 0.000057 & -0.000094 & 0.000155 & 0.000078 \\
    3 & 0.000109 & 0.000162 & -0.000010 & 0.000094 & -0.000078 & -0.000025 \\
    4 & -0.000090 & -0.000204 & 0.000169 & -0.000055 & 0.000315 & 0.000199 \\
    \bottomrule
    \end{tabular}
    
    \end{subtable}%

    \caption{\textbf{Magnetization} Single emulation}
    \label{tab:hubbard_large_single_emulation_magnetizations}
\end{table}

\clearpage
\bibliographystyle{unsrt}
\bibliography{refs}

\end{document}